\documentclass[aps, prd, amsmath, preprintnumbers, floats, floatfix, twocolumn, nofootinbib, showpacs]{revtex4-1}
\usepackage{graphicx}
\usepackage{color}

\newcommand{\bi}{\bibitem}
\newcommand{\be}{\begin{eqnarray}}
\newcommand{\ee}{\end{eqnarray}}

\begin{document}

\title{Testing the space-time geometry around black hole candidates with the analysis of the broad K$\alpha$ iron line}

\author{Cosimo Bambi}
\email{bambi@fudan.edu.cn}
\affiliation{Center for Field Theory and Particle Physics \& Department of Physics,\\ 
Fudan University, 200433 Shanghai, China}
\affiliation{Arnold Sommerfeld Center for Theoretical Physics,\\
Ludwig-Maximilians-Universit\"at M\"unchen, D-80333 Munich, Germany}

\date{\today}


\begin{abstract}
Astrophysical black hole candidates are thought to be the Kerr black holes
predicted by General Relativity, but there is not yet clear evidence that 
the geometry of the space-time around these objects is really described by 
the Kerr metric. In order to confirm the Kerr black hole hypothesis, we have 
to observe strong gravity features and check that they are in agreement with the 
ones predicted by General Relativity. In this paper, I study the broad K$\alpha$ 
iron line, which is often seen in the X-ray spectrum of both stellar-mass and 
super-massive black hole candidates and whose shape is supposed to be 
strongly affected by the space-time geometry. As found in previous studies 
in the literature, there is a strong correlation between the spin parameter
and the deformation parameter; that is, the line emitted around a Kerr black 
hole with a certain spin can be very similar to the one coming from the 
space-time around a non-Kerr object with a quite different spin. Despite that, 
the analysis of the broad K$\alpha$ iron line is potentially more powerful than 
the continuum-fitting method, as it can put an interesting bound on possible deviations 
from the Kerr geometry independently of the value of the spin parameter 
and without additional measurements.
\end{abstract}


\maketitle


\section{Introduction}

Today we think the final product of the gravitational collapse is a black hole 
(BH) and we know several strong astrophysical candidates~\cite{bh-r}. In 
4-dimensional General Relativity, an uncharged BH is described by the Kerr 
solution and it is completely characterized by two quantities, the mass $M$ 
and the spin parameter $a_* = J/M^2$, where $J$ is the BH spin angular 
momentum~\cite{kerr}\footnote{Throughout the paper, I use units in which $G_{\rm N} 
= c = 1$, unless stated otherwise.}. A fundamental limit for a 4-dimensional Kerr BH 
is the bound $|a_*| \le 1$, which is the condition for the existence of the event 
horizon. For $|a_*| > 1$, there is no horizon and the Kerr metric describes
the gravitational field of a naked singularity, which is forbidden by the Weak
Cosmic Censorship Conjecture~\cite{wccc}.

The study of the orbital motion of individual stars around a BH candidate can 
provide robust measurements of the mass of the latter: these stars are typically far 
from the compact object and one can estimate $M$ by using Newtonian mechanics, 
with no assumptions about the nature of the BH candidate. On the contrary, the 
measurement of the spin is much more challenging. The spin has no effects in 
Newtonian gravity and it can be estimated only probing the space-time close
to the compact object. That can be potentially achieved by studying the properties
of the electromagnetic radiation emitted by the gas of the accretion disk. The
two most popular techniques to try to estimate $a_*$ are the continuum-fitting
method~\cite{cfm-r} and the analysis of the broad K$\alpha$ iron line~\cite{fe-r}.
In both the approaches, an important assumption is that the inner edge of the disk
is at the innermost stable circular orbit (ISCO). Since in the Kerr background there 
is a one-to-one correspondence between the ISCO radius and the spin parameter 
$a_*$, fitting the data we can get an estimate of $a_*$, supposing that all the 
astrophysical effects are well understood and properly taken into account.

As there is not yet clear evidence that the geometry of the space-time around
BH candidates is the one described by the Kerr solution, different authors
have proposed different ways to test this hypothesis. The first proposal was
put forward by Ryan, who suggested observing the gravitational waves
emitted by an extreme-mass ratio inspiral (EMRI), i.e. a system consisting of
a stellar-mass compact object orbiting around a super-massive BH candidate~\cite{ryan}:
as future space-based gravitational wave detectors will be able to observe
$\sim 10^4 - 10^6$ gravitational wave cycles emitted by an EMRI while the 
stellar-mass body is inspiraling into the gravitational field of the super-massive 
object, even a small deviation from the Kerr geometry will build up an observable
dephasing in the gravitational waveforms, thus allowing one to map the 
space-time of the super-massive BH candidate with very high accuracy~\cite{emri}.
Besides tests based on gravitational waves, the space-time geometry
around BH candidates can be probed even by very accurate observations
of the orbital motion of stars. In the case of a stellar-mass BH candidate in
an X-ray binary system, that is possible if the companion star is a radio
pulsar~\cite{wex}. In the case of the super-massive BH candidate at the
center of our Galaxy, that might be achieved by monitoring stars orbiting at 
milliparsec distances from the compact object~\cite{will}.

Recently, there has been new interest in the 
subject~\cite{cfm-b1,cb-apj,cfm-b2,cb,spin,bb2,jp-fe,jp-em,cn1,cn2,krawcz,vlbi,kono}, 
especially thanks to significant progress in the understanding of the properties of the 
electromagnetic radiation emitted by the gas in the accretion disk and to 
near future high-resolution sub-millimeter VLBI experiments. 
It has been shown that the continuum-fitting method~\cite{cfm-b1,cb-apj} and
the analysis of the broad K$\alpha$ iron line~\cite{jp-fe} can be easily
generalized to non-Kerr space-times and be used to test the Kerr-nature
of astrophysical BH candidates. Unlike other proposals, the two
techniques can be applied to already available X-ray data and therefore
they are both of particular interest. The continuum-fitting method is
based on the analysis of the thermal spectrum of a geometrically thin
accretion disk and it can be applied only to stellar-mass BH candidates
-- in the case of super-massive BH candidates, the thermal spectrum 
falls in the UV range and dust absorption makes accurate measurements
impossible. However, the fit of the disk's thermal spectrum actually
measures the radiative efficiency of the Novikov-Thorne model,
$\eta = 1 - E_{\rm ISCO}$, where $E_{\rm ISCO}$ is the specific energy
of a test-particle at the ISCO radius~\cite{cfm-b1,cfm-b2}. In other words,
the thermal spectrum of the disk around a Kerr BH with spin $a_*$ and
radiative efficiency $\eta$ is extremely similar, and eventually 
indistinguishable, from the one around a non-Kerr object with different
spin parameter but same radiative efficiency.

The analysis of the broad K$\alpha$ iron line is based on some more subtle
assumptions, but it is thought to be potentially a more powerful approach 
to probe the space-time geometry around BH candidates. A preliminary
study for using the broad K$\alpha$ iron line to test the nature of
astrophysical BH candidates has been presented in Ref.~\cite{jp-fe}.
The aim of the present paper is to extend that work. I compare the K$\alpha$ 
iron line produced around a Kerr BH with the one generated around more generic 
objects. As already found in Ref.~\cite{jp-fe}, there is a strong correlation between
the spin and the deformation parameter. I consider several specific cases,
covering a wide range of spins and inclination angles. The results are
reported in Tab.~\ref{tab1}. It is also interesting to note that even a quite
deformed object can provide a K$\alpha$ line similar to the one expected
in a Kerr background, as is shown by the reduced $\chi^2$ in the top 
right panel of Fig.~\ref{f-nk}. Despite that, the technique can still put an
interesting bound on possible deviations from the Kerr geometry, 
independently of the value of $a_*$ and without additional measurements, 
which, in general, is not the case for the continuum-fitting method.

The content of the present paper is as follows. In Section~\ref{s-k}, I
review the analysis of the K$\alpha$ iron line in the Kerr background.
In Section~\ref{s-jp}, I consider the non-Kerr metric proposed in~\cite{jp-m}
and I extend the analysis of Ref.~\cite{jp-fe} to use the K$\alpha$ line to
constrain the space-time geometry around an astrophysical BH candidate.
I cover a wide range of spins and inclinations.
In Section~\ref{s-e}, I discuss the results of Section~\ref{s-jp}, pointing
out the differences with the continuum-fitting method and showing 
the outcome of a possible combination of the two techniques. Summary 
and conclusions are in Section~\ref{s-c}.

\begin{figure*}
\begin{center}
\includegraphics[type=pdf,ext=.pdf,read=.pdf,width=8.5cm]{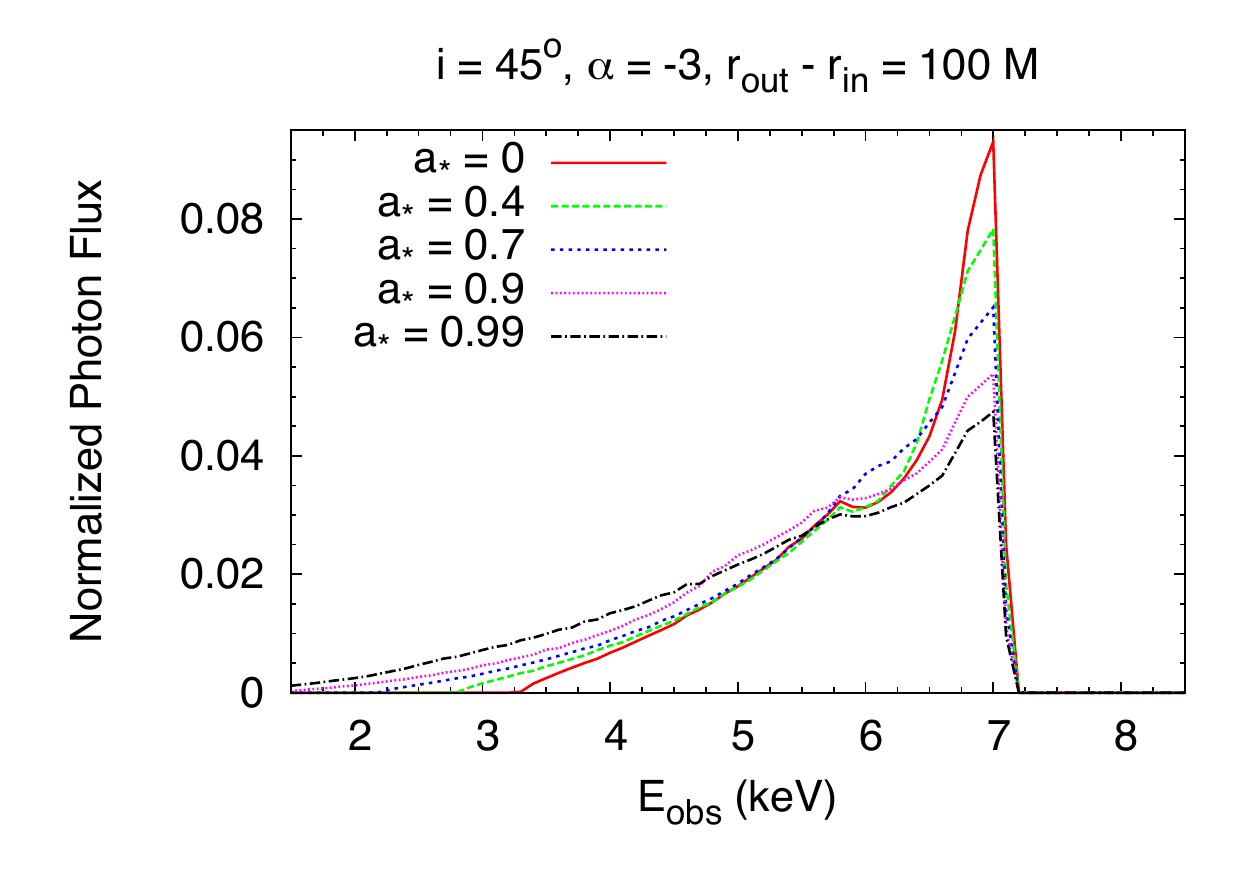}
\includegraphics[type=pdf,ext=.pdf,read=.pdf,width=8.5cm]{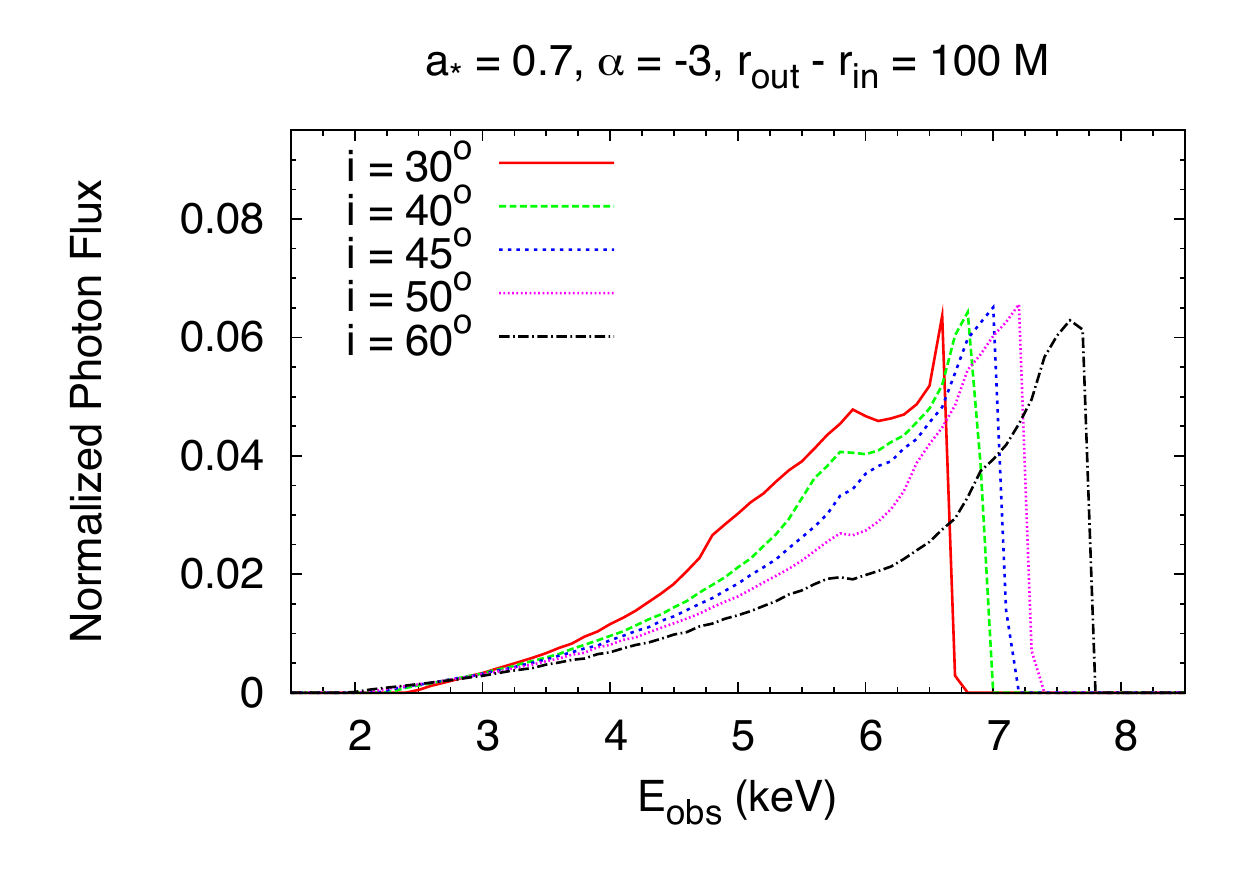} \\
\includegraphics[type=pdf,ext=.pdf,read=.pdf,width=8.5cm]{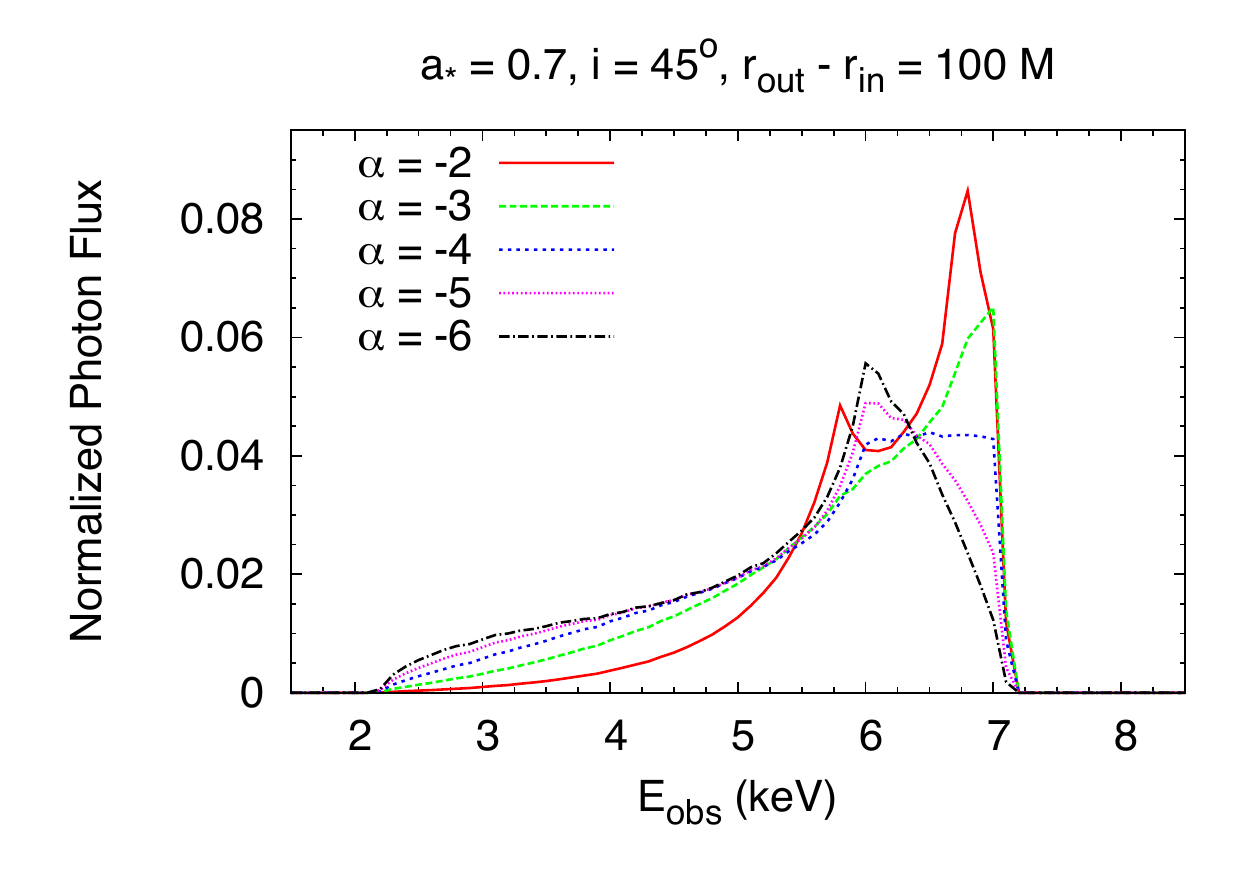}
\includegraphics[type=pdf,ext=.pdf,read=.pdf,width=8.5cm]{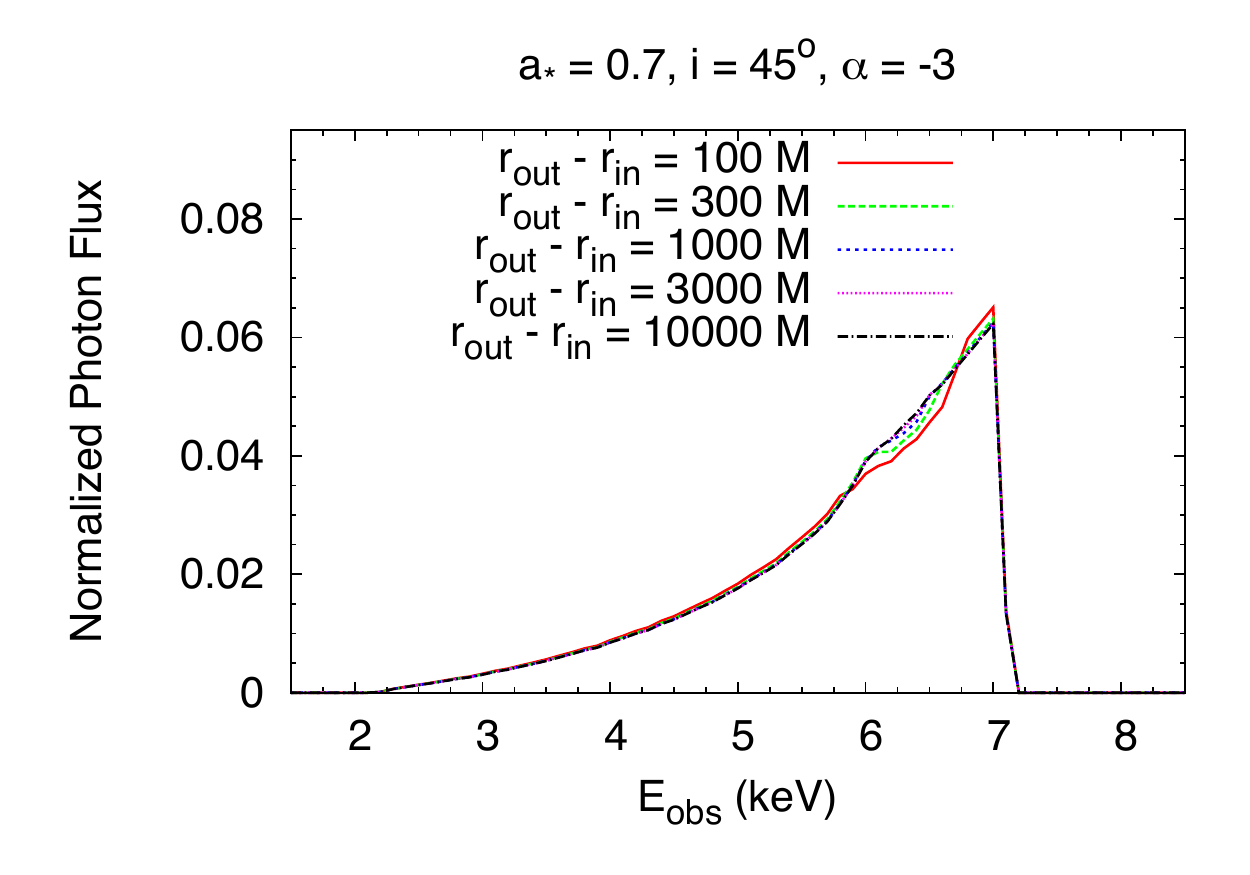}
\vspace{-0.5cm}
\caption{Broad K$\alpha$ iron line in the Kerr background for
different values of the parameters of the model. {\it Top left panel:} effect
of the spin parameter $a_*$ for a viewing angle $i = 45^\circ$, intensity 
profile with index $\alpha = -3$, and emissivity region $r_{\rm out} - r_{\rm in} 
= 100 \; M$. {\it Top right panel:} effect of the viewing angle for a spin 
parameter $a_* = 0.7$, intensity profile with index $\alpha = -3$, and 
emissivity region $r_{\rm out} - r_{\rm in} = 100 \; M$. {\it Bottom left
panel:} effect of the intensity profile index $\alpha$, for a spin parameter
$a_* = 0.7$, viewing angle $i = 45^\circ$, and emissivity region 
$r_{\rm out} - r_{\rm in} = 100 \; M$. {\it Bottom right panel:} effect of the
emissivity region for a spin parameter $a_* = 0.7$, viewing angle 
$i = 45^\circ$, and intensity profile with index $\alpha = -3$. See the text 
for details.}
\label{f-kerr}
\end{center}
\end{figure*}

\section{Broad K$\alpha$ iron line in the Kerr background \label{s-k}}

The X-ray spectrum of both stellar-mass and super-massive BH candidates
is usually characterized by the presence of a power-law component. This
feature is commonly interpreted as the inverse Compton scattering of 
thermal photons by electrons in a hot corona above the accretion disk.
The geometry of the corona is not known and several models have been
proposed. Such a ``primary component'' irradiates
also the accretion disk, producing a ``reflection component'' in the X-ray 
spectrum. The illumination of the cold disk by the primary component
also produces spectral lines by fluorescence. The strongest line is the 
K$\alpha$ iron line at 6.4~keV. This line is intrinsically narrow in frequency,
while the one observed appears broadened and skewed. The interpretation 
is that the line is strongly altered by special and general relativistic effects, 
which produce a characteristic profile first predicted in Ref.~\cite{fab89} and 
then observed for the first time in the ASCA data of the Seyfert~1 galaxy 
MCG-6-30-15~\cite{tan95}. In the specific case of MCG-6-30-15, this line
is extraordinarily stable, in spite of a substantial variability of the continuum,
suggesting that the analysis of its shape can be used to probe the
geometry of the space-time around the BH candidate. It should be borne 
in mind, however, that the relativistic origin of the observed broad K$\alpha$ 
iron lines is not universally accepted, and some authors have proposed 
different explanations~\cite{tita}.

Within the interpretation of a relativistically broadened K$\alpha$ iron line,
the shape of the line is primarily determined by the background metric, the 
geometry of the emitting region, the disk emissivity, and the disk's inclination 
angle with respect to the line of sight of the distant observer~\cite{fab89}. In the 
standard framework of a Kerr background, $M$ sets the length of the system, so
everything scales as $M$ or as some power of $M$, without affecting the 
shape of the line. The only relevant parameter of the background geometry is
thus the spin $a_*$. In those sources for which there is indication that the
K$\alpha$ iron line is mainly emitted close to the compact object, the
emission region may be thought to range from the ISCO radius, 
$r_{\rm in} = r_{\rm ISCO}$, to some outer radius $r_{\rm out}$. 
However, even more complicated geometries have been
proposed, see e.g. Ref.~\cite{er}. In principle, the disk emissivity may be
theoretically calculated. In practice, that is not feasible at present.
The simplest choice is an intensity profile $I_{\rm e} \propto r^{\alpha}$,
with $\alpha < 0$ a free parameter to be determined during the fitting 
procedure. The fourth parameter is the inclination of the disk with respect
to the line of sight of the distant observer, $i$. The dependence of the Kerr
iron line profile on $a_*$, $i$, $\alpha$, and $r_{\rm out}$ and summarized 
in this section has been analyzed in detail by many authors,
starting with Ref.~\cite{fab89}.

In what follows, I use the ray-tracing code described in~\cite{cb-apj}. 
The photon flux number density as measured by a distant 
observer is given by
\be
N_{E_{\rm obs}} &=& \frac{1}{E_{\rm obs}} 
\int I_{\rm obs}(E_{\rm obs}) d \Omega_{\rm obs} = \nonumber\\ 
&=& \frac{1}{E_{\rm obs}} \int g^3 I_{\rm e}(E_{\rm e}) 
d \Omega_{\rm obs} \, .
\ee
where $I_{\rm obs}$ and $E_{\rm obs}$ are, respectively, the 
specific intensity of the radiation and the photon energy as measured by 
the distant observer, $d \Omega_{\rm obs}$ is the element of the solid 
angle subtended by the image of the disk on the observer's sky,
$I_{\rm e}$ and $E_{\rm e}$ are, respectively, the local specific intensity
of the radiation and the photon energy in the rest frame of the
emitter, and $g = E_{\rm obs}/E_{\rm e}$ is the redshift factor.
$I_{\rm obs} = g^3 I_{\rm e}$ follows from the Liouville's theorem.
Unlike Ref.~\cite{cb-apj}, here I assume that the disk emission is
monochromatic (the rest frame energy is $E_{\rm{K}\alpha} = 6.4$~keV)
and isotropic with a power-law radial profile:
\be
I_{\rm e}(E_{\rm e}) \propto \delta (E_{\rm e} - E_{\rm{K}\alpha}) r^{\alpha} \, .
\ee

The resulting broad K$\alpha$ iron lines for different values of the 
model parameters (spin $a_*$, inclination angle $i$, power-law index 
$\alpha$, and outer radius $r_{\rm out}$) are shown in Fig.~\ref{f-kerr}.
The photon flux has been normalized so that
\be
\int N_{E_{\rm obs}} d E_{\rm obs} = {\rm constant} \, ,
\ee
as only the shape matters.
The spin (left top panel) can be determined by the low-energy tail:
a higher spin implies an inner radius closer to the compact object and 
therefore a larger fraction of photon is affected by a strong gravitational 
redshift ($r_{\rm ISCO}/M = 6$ for $a_* = 0$ and monotonically goes to 1 
for $a_* = 1$). 
The disk's inclination angle $i$ (right top panel) moves the blue-shifted part
of the spectrum: for small inclination angles, the Doppler boosting is 
not relevant, while for large inclination angles it becomes
more and more important. The power-law index $\alpha$ (left bottom 
panel) balances the importance of the emission in the innermost 
region with respect to the one at larger radii. For high values of 
$\alpha$, e.g. $\alpha = -2$, the contribution of the emission at
relatively large radii is significant, so the Doppler blue-shifted part of
the spectrum results in a prominent peak, while the gravitational red-shifted part has
a lower flux. For lower and lower values of $\alpha$, the photon
flux in the Doppler blue-shifted part of the spectrum decreases,
while the one in the gravitational red-shifted part increases.
Lastly, the position of the outer radius of the emission region $r_{\rm out}$
(right bottom panel) changes the position of the low energy peak of the
spectrum, although the effect is small for $\alpha = -3$ or lower.

\begin{figure*}
\begin{center}
\includegraphics[type=pdf,ext=.pdf,read=.pdf,width=8.5cm]{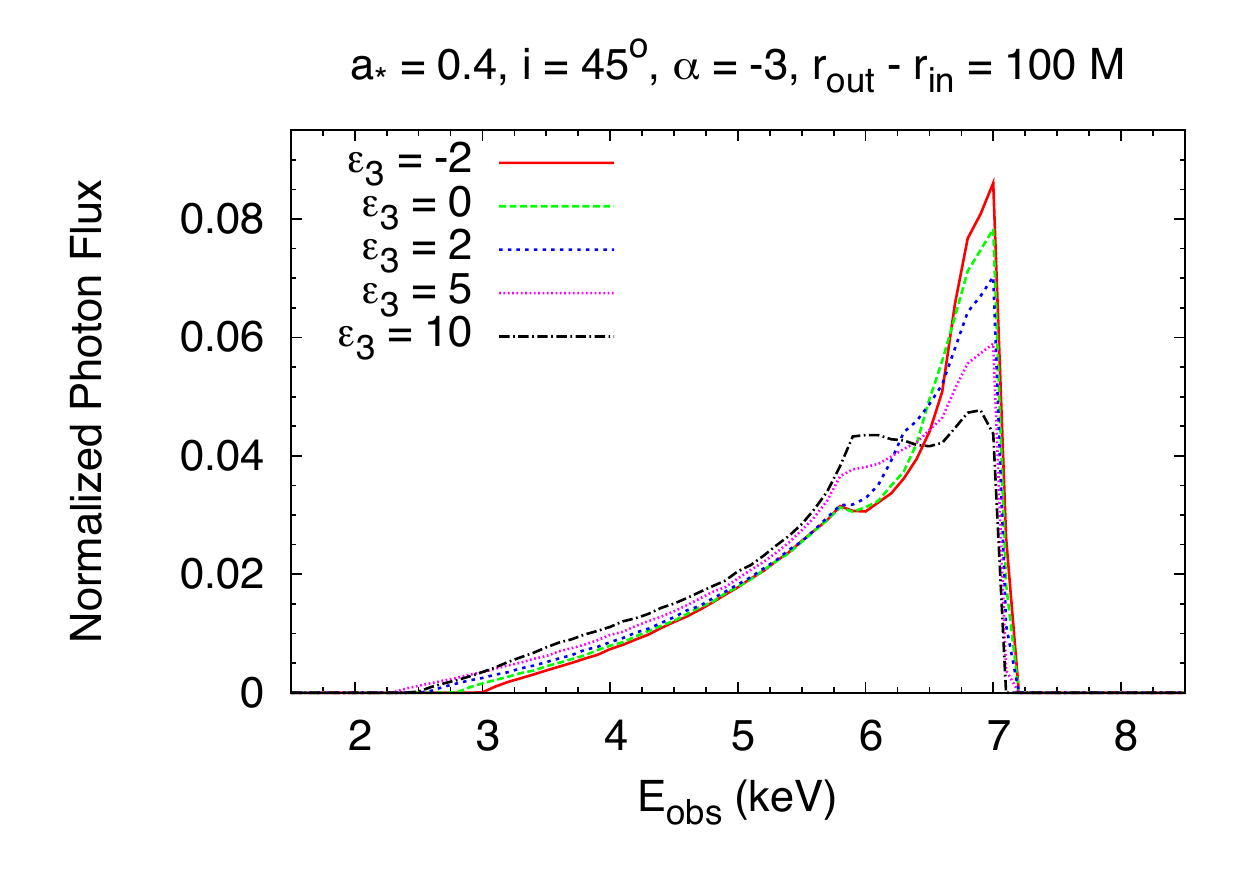}
\includegraphics[type=pdf,ext=.pdf,read=.pdf,width=8.5cm]{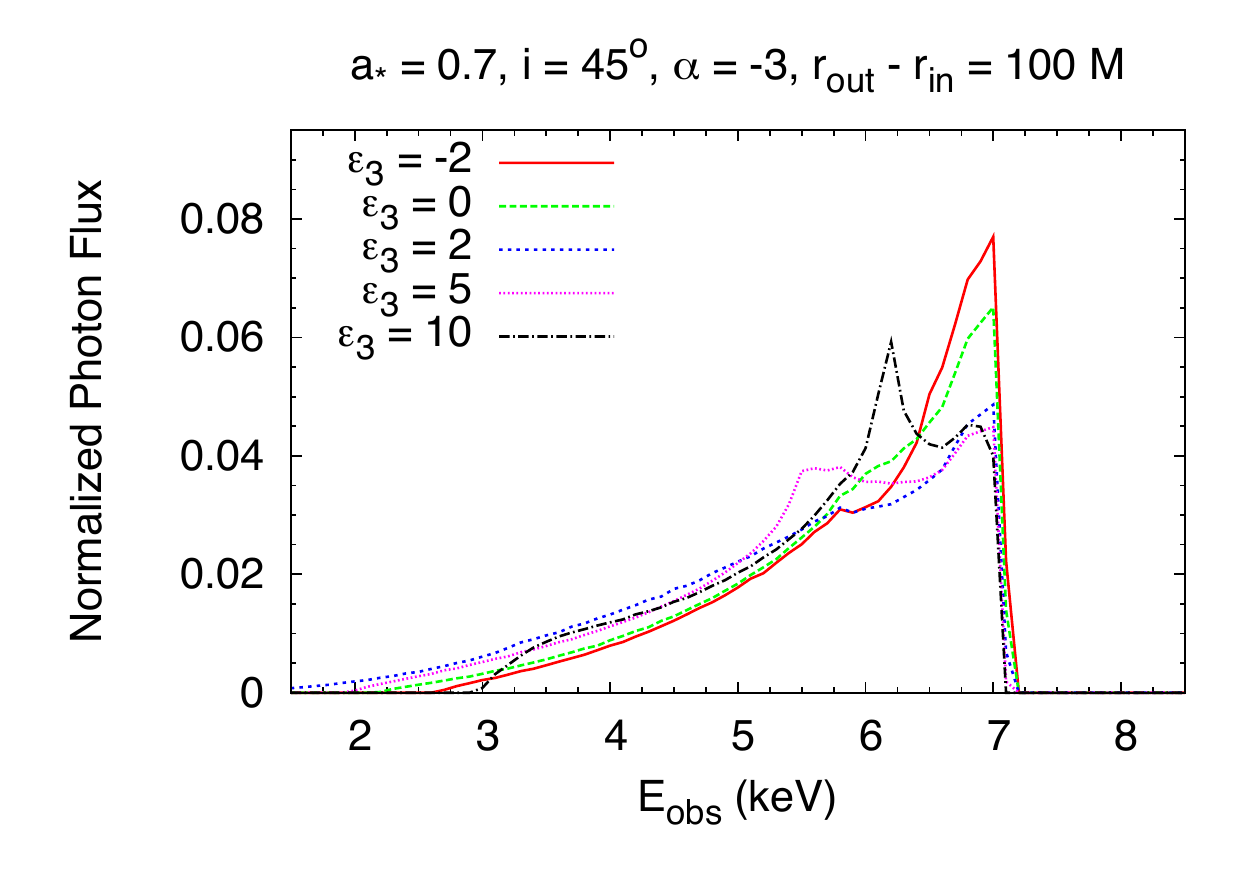}
\vspace{-0.5cm}
\caption{Broad K$\alpha$ iron line in the JP background for different 
values of the deformation parameter $\epsilon_3$. {\it Left panel:} spin 
parameter $a_* = 0.4$. {\it Right panel:} spin parameter $a_* = 0.7$.
In both the panels, the viewing angle is $i = 45^\circ$, the intensity profile
index is $\alpha = -3$, and the emissivity region is $r_{\rm out} - r_{\rm in} 
= 100 \; M$. See the text for details.}
\label{f-jp}
\end{center}
\end{figure*}

\begin{figure*}
\begin{center}
\vspace{-2.5cm}
\includegraphics[type=pdf,ext=.pdf,read=.pdf,width=8cm]{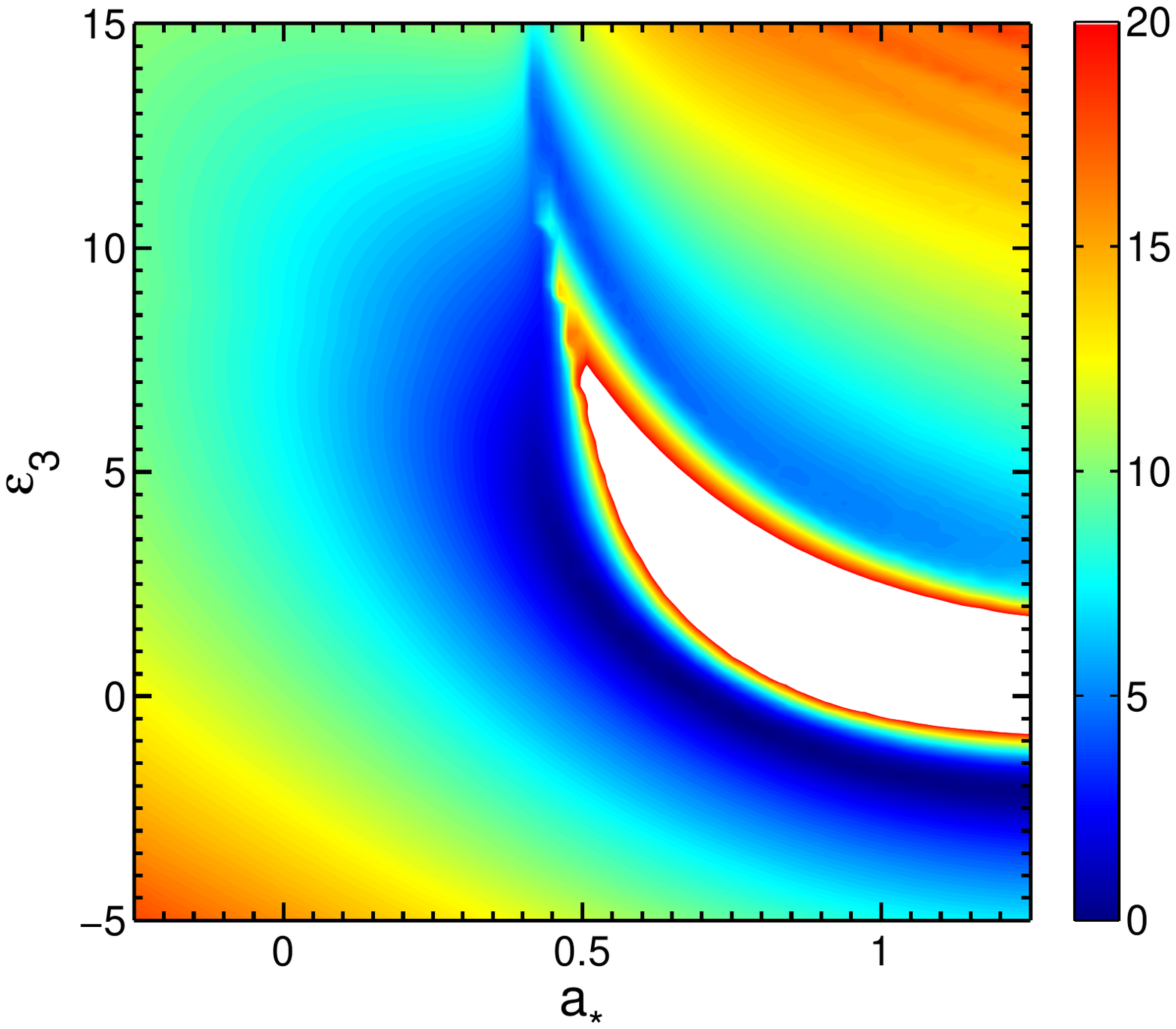}
\includegraphics[type=pdf,ext=.pdf,read=.pdf,width=8cm]{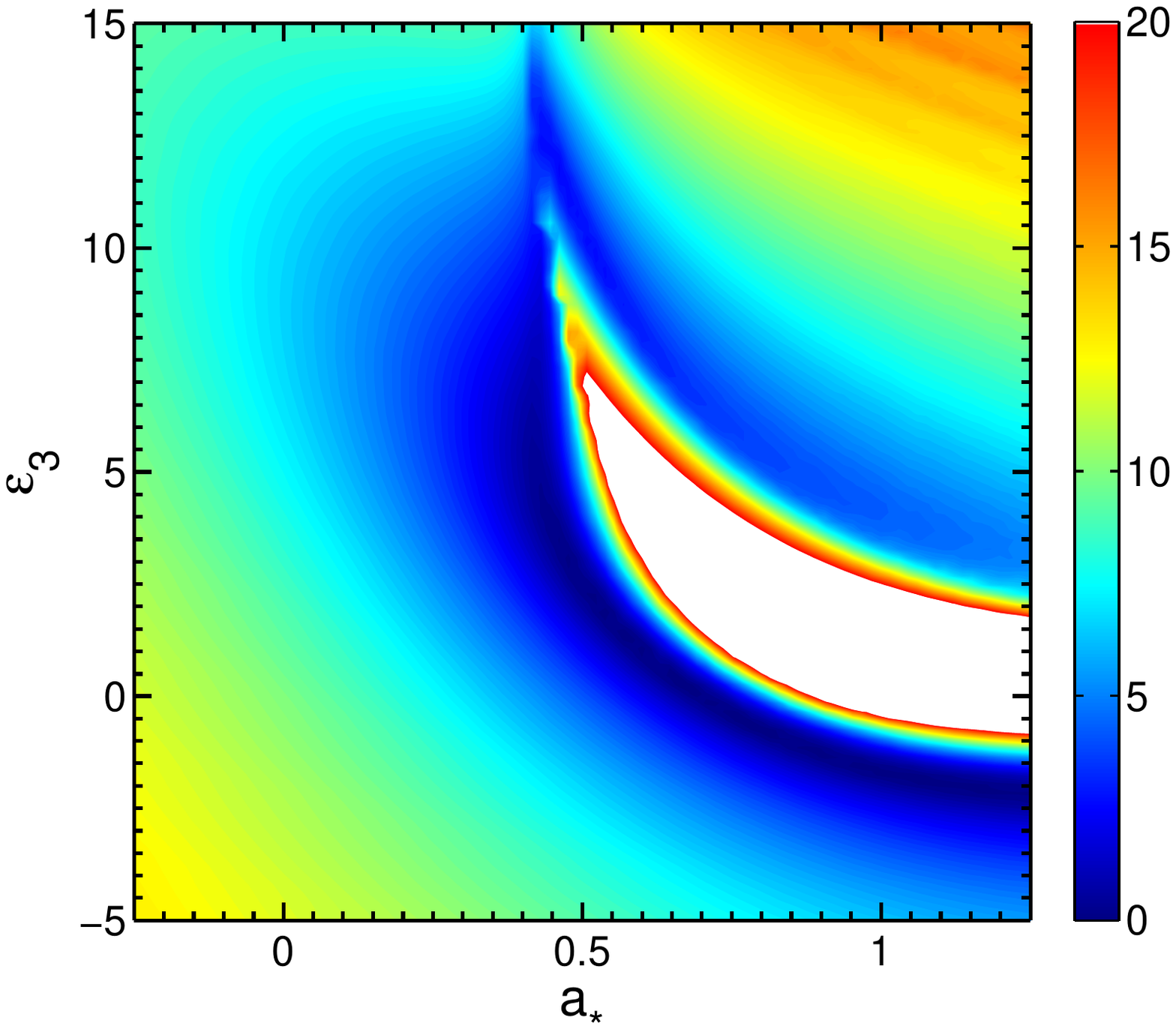} \\
\vspace{-5cm}
\includegraphics[type=pdf,ext=.pdf,read=.pdf,width=8cm]{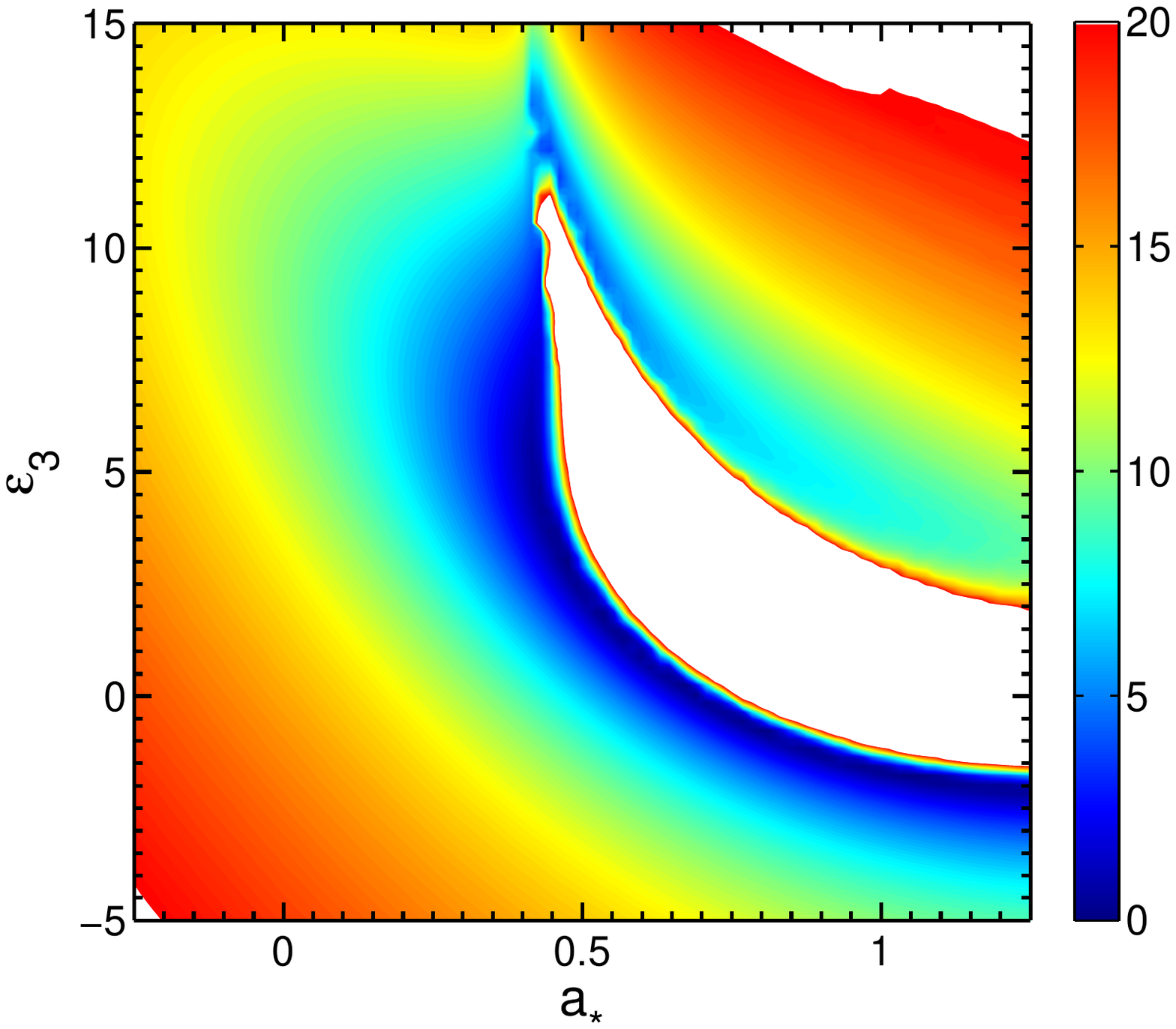}
\includegraphics[type=pdf,ext=.pdf,read=.pdf,width=8cm]{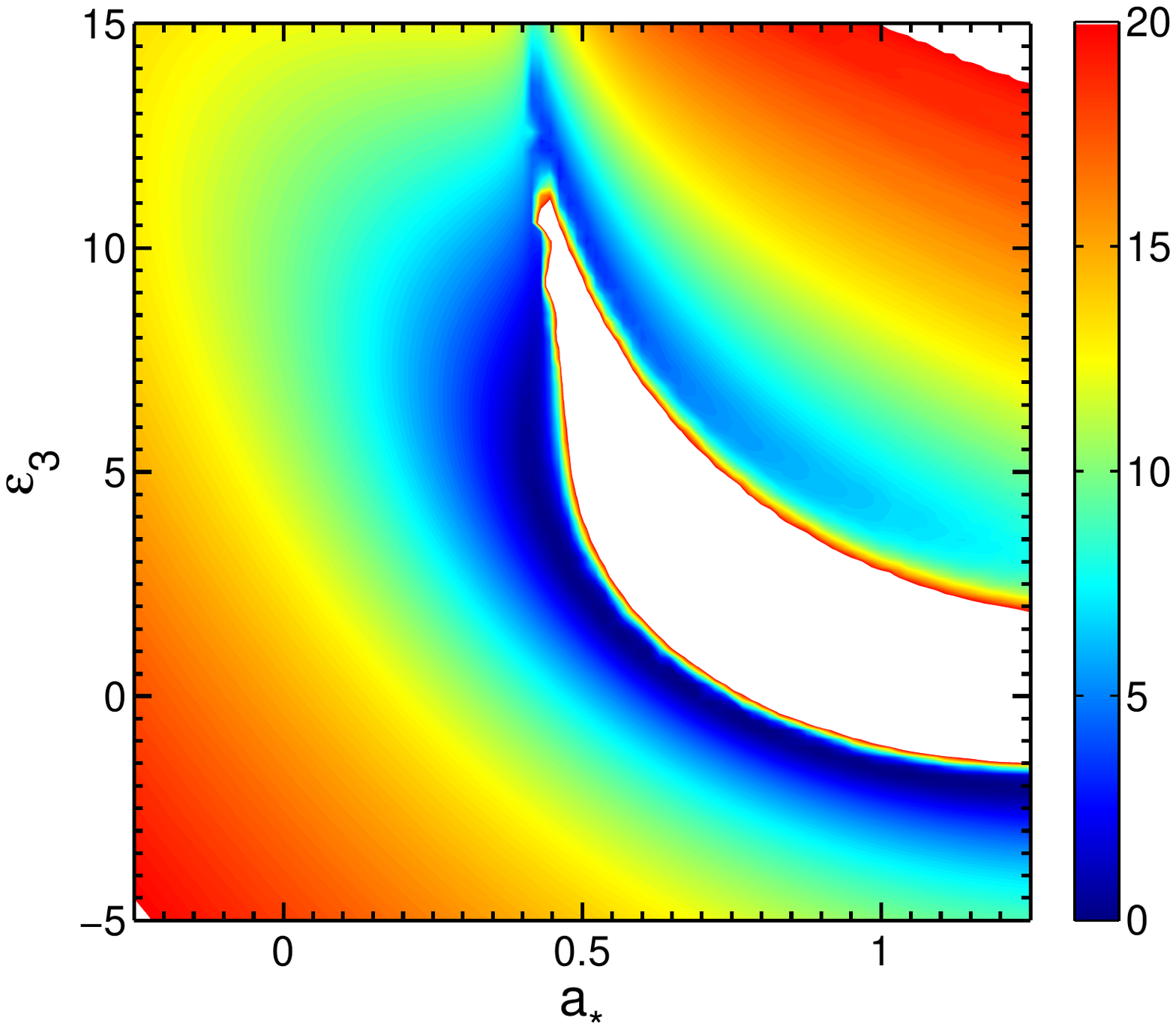} \\
\vspace{-5cm}
\includegraphics[type=pdf,ext=.pdf,read=.pdf,width=8cm]{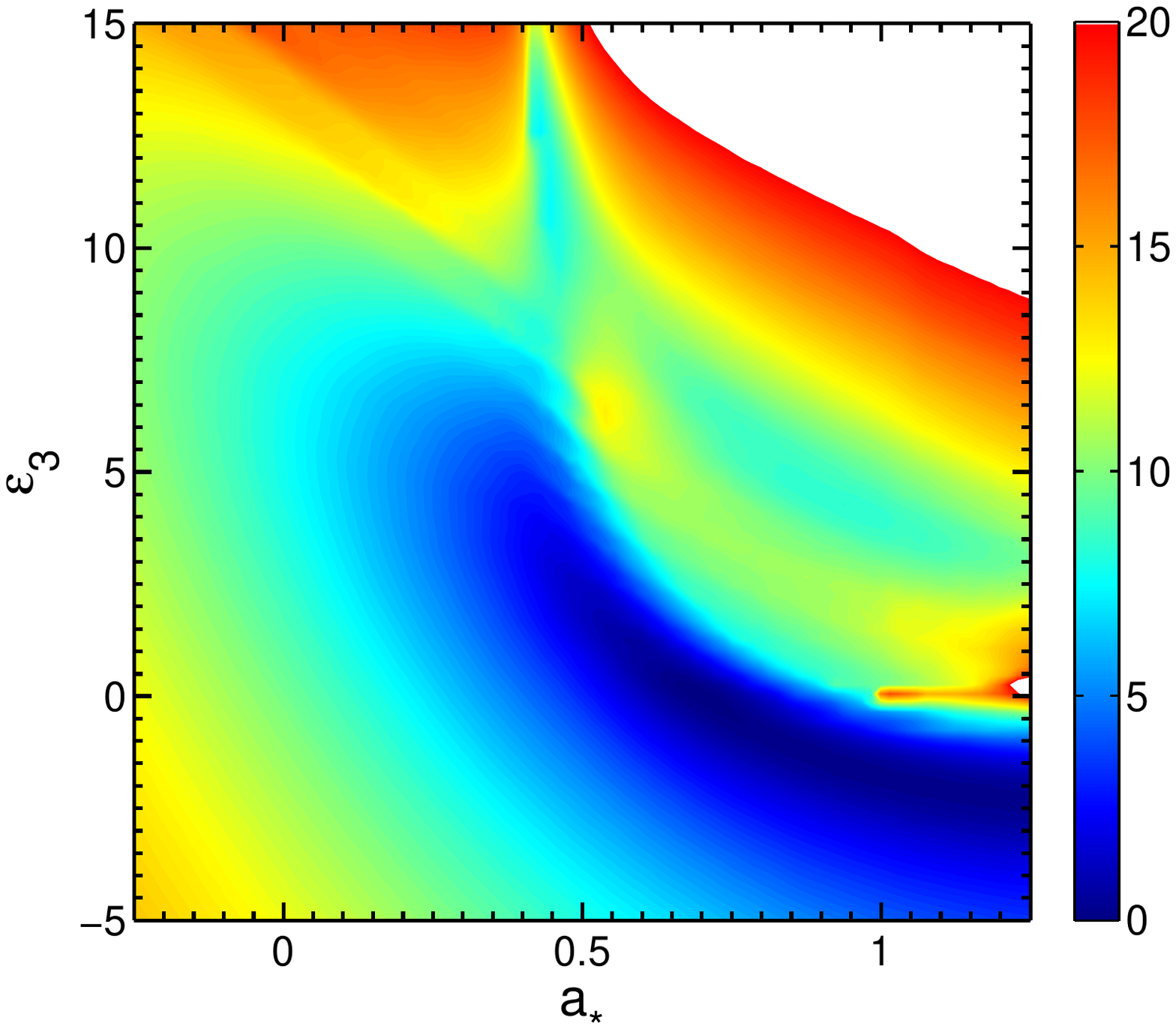}
\includegraphics[type=pdf,ext=.pdf,read=.pdf,width=8cm]{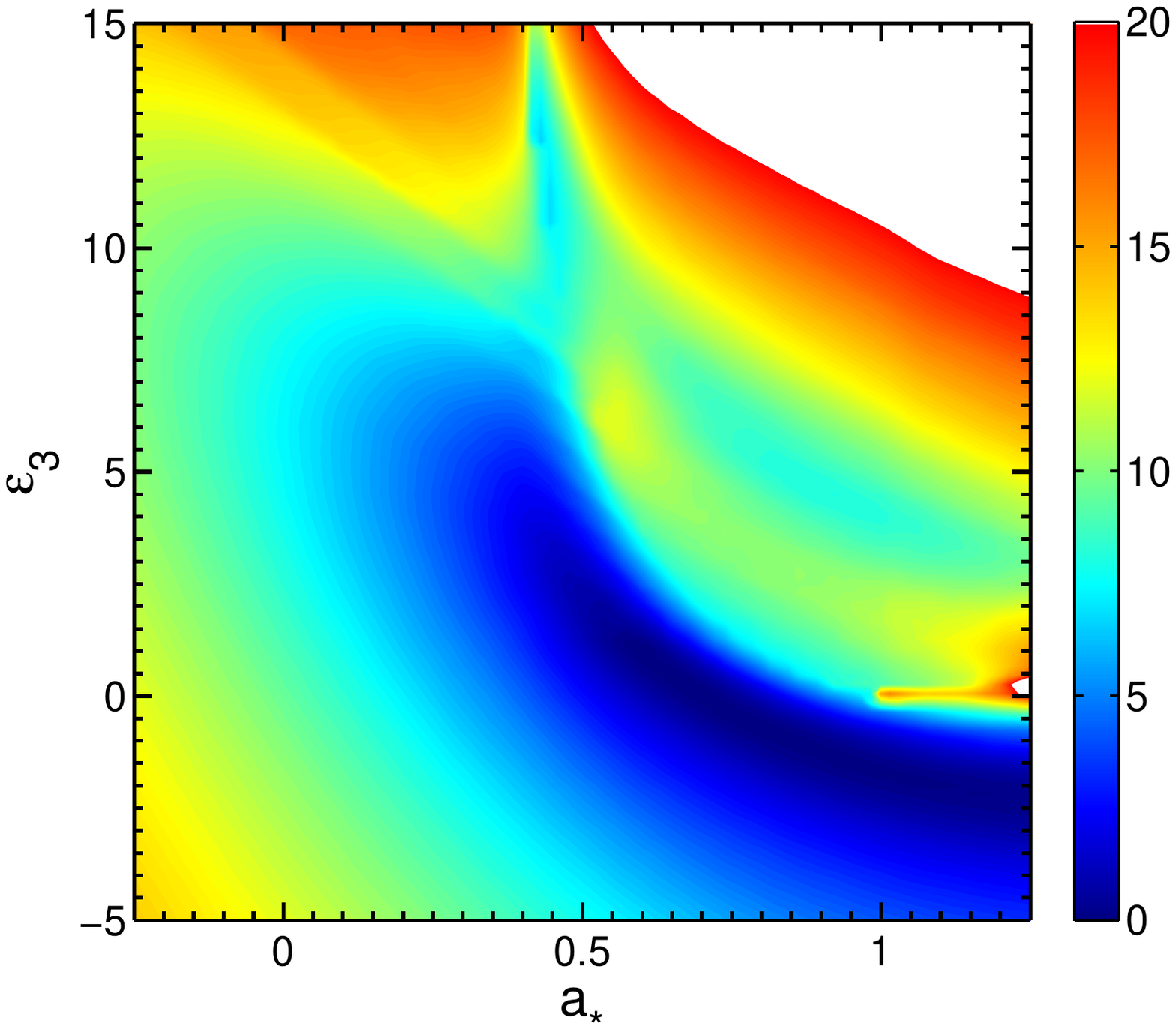}
\vspace{-2.5cm}
\caption{$\chi^2_{\rm red}$ from the comparison of the broad K$\alpha$ iron 
line produced in a Kerr space-time with spin parameter $\tilde{a}_* = 0.70$ and 
the one in a JP space-time with spin parameter $a_*$ and deformation parameter 
$\epsilon_3$. {\it Top left panel:} $\tilde{i}=i=45^\circ$. {\it Top right panel:} 
$\tilde{i}=45^\circ$ and $i$ free. {\it Central left panel:} $\tilde{i}=i=15^\circ$.
{\it Central right panel:} $\tilde{i}=15^\circ$ and $i$ free. {\it Bottom left panel:} 
$\tilde{i}=i=75^\circ$. {\it Bottom right panel:} $\tilde{i}=75^\circ$ and $i$ free.
The other parameters of the model are $\tilde{\alpha}=\alpha=-3$ and 
$\tilde{r}_{\rm out} - \tilde{r}_{\rm in}=r_{\rm out}- r_{\rm in} =100$~$M$.
See the text for details.}
\label{f-chi2-1}
\end{center}
\end{figure*}

\begin{figure*}
\begin{center}
\vspace{-2.5cm}
\includegraphics[type=pdf,ext=.pdf,read=.pdf,width=8cm]{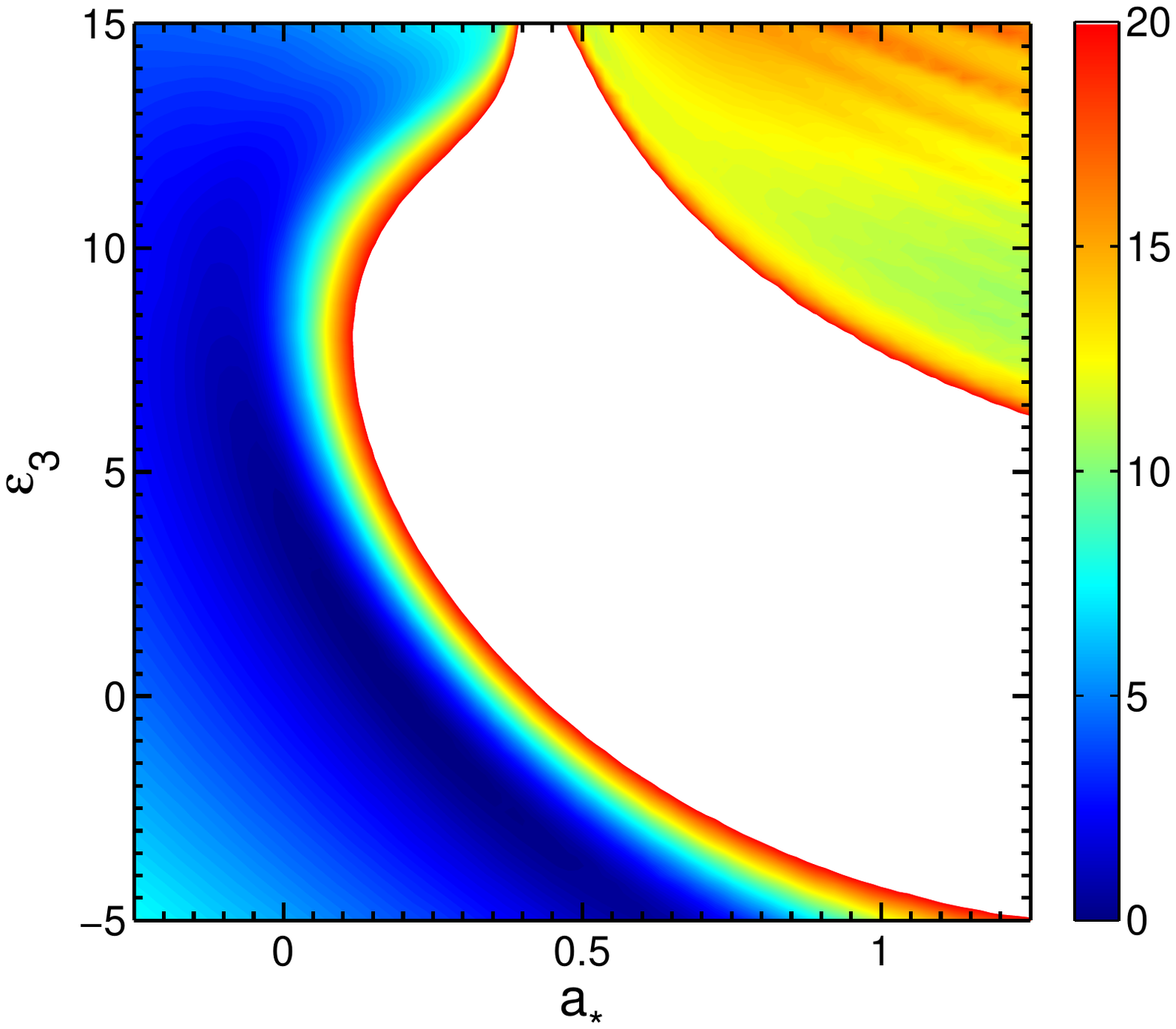}
\includegraphics[type=pdf,ext=.pdf,read=.pdf,width=8cm]{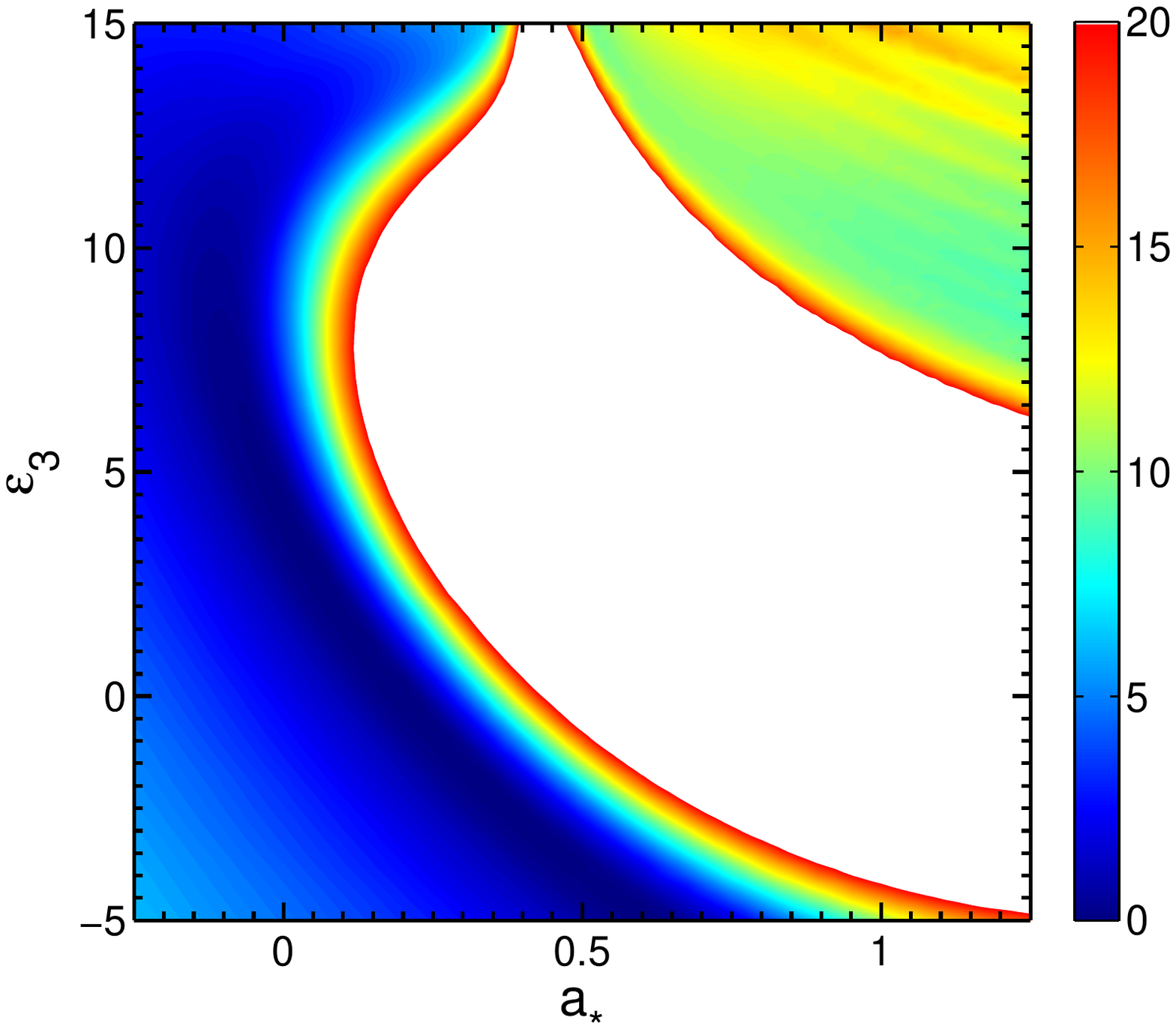} \\
\vspace{-5cm}
\includegraphics[type=pdf,ext=.pdf,read=.pdf,width=8cm]{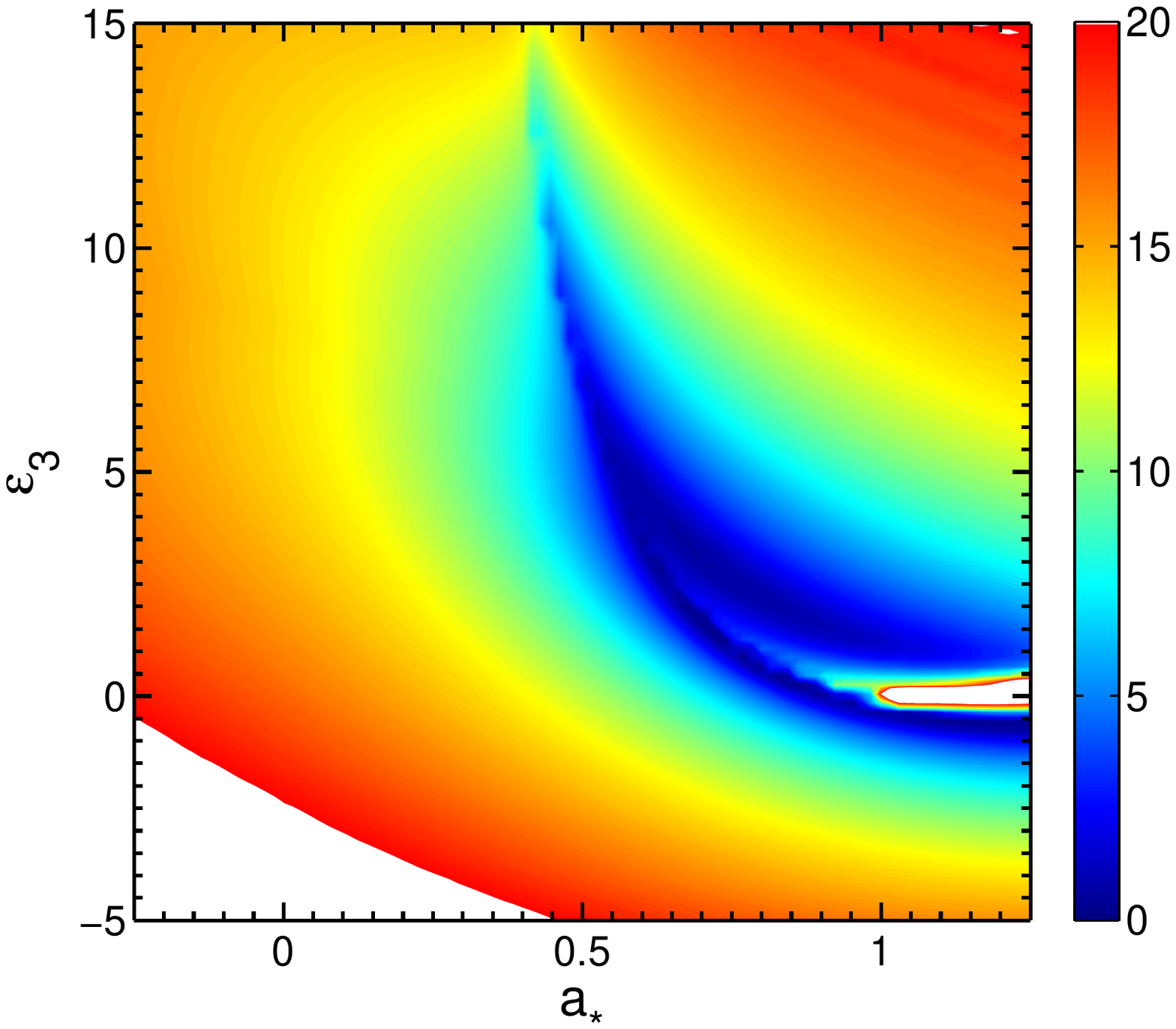}
\includegraphics[type=pdf,ext=.pdf,read=.pdf,width=8cm]{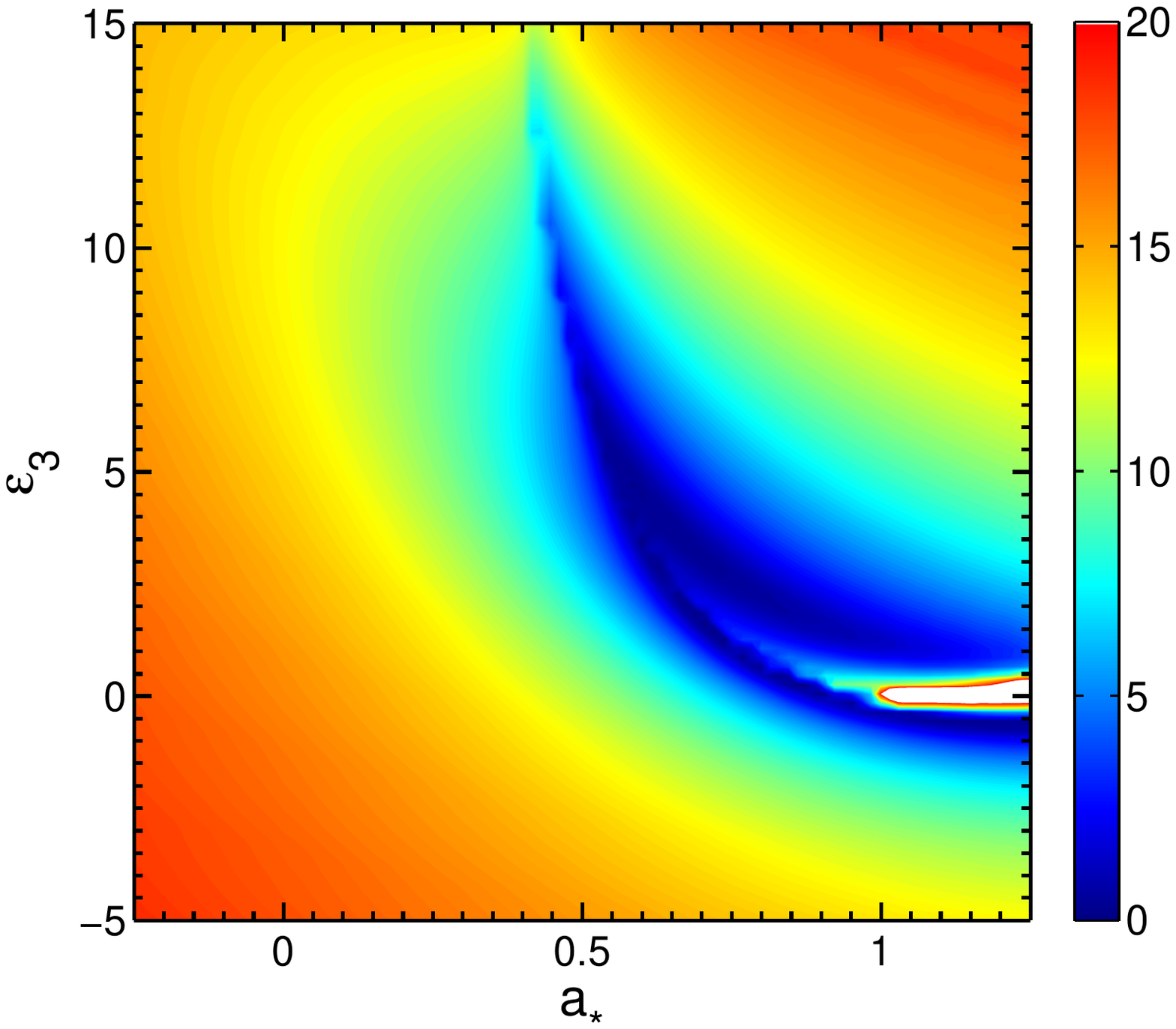}
\vspace{-2.5cm}
\caption{$\chi^2_{\rm red}$ from the comparison of the broad K$\alpha$ iron 
line produced in a Kerr space-time with spin parameter $\tilde{a}$ and 
the one in a JP space-time with spin parameter $a_*$ and deformation parameter 
$\epsilon_3$. {\it Top left panel:} $\tilde{a}_* = 0.20$ and $\tilde{i}=i=45^\circ$. 
{\it Top right panel:} $\tilde{a}_* = 0.20$, $\tilde{i}=45^\circ$, and $i$ free. 
{\it Bottom left panel:} $\tilde{a}_* = 0.92$ and $\tilde{i}=i=45^\circ$. 
{\it Bottom right panel:} $\tilde{a}_* = 0.92$, $\tilde{i}=45^\circ$, and $i$ free.
See the text for details.}
\label{f-chi2-2}
\end{center}
\end{figure*}

\begin{figure*}
\begin{center}
\vspace{-2.5cm}
\includegraphics[type=pdf,ext=.pdf,read=.pdf,width=8cm]{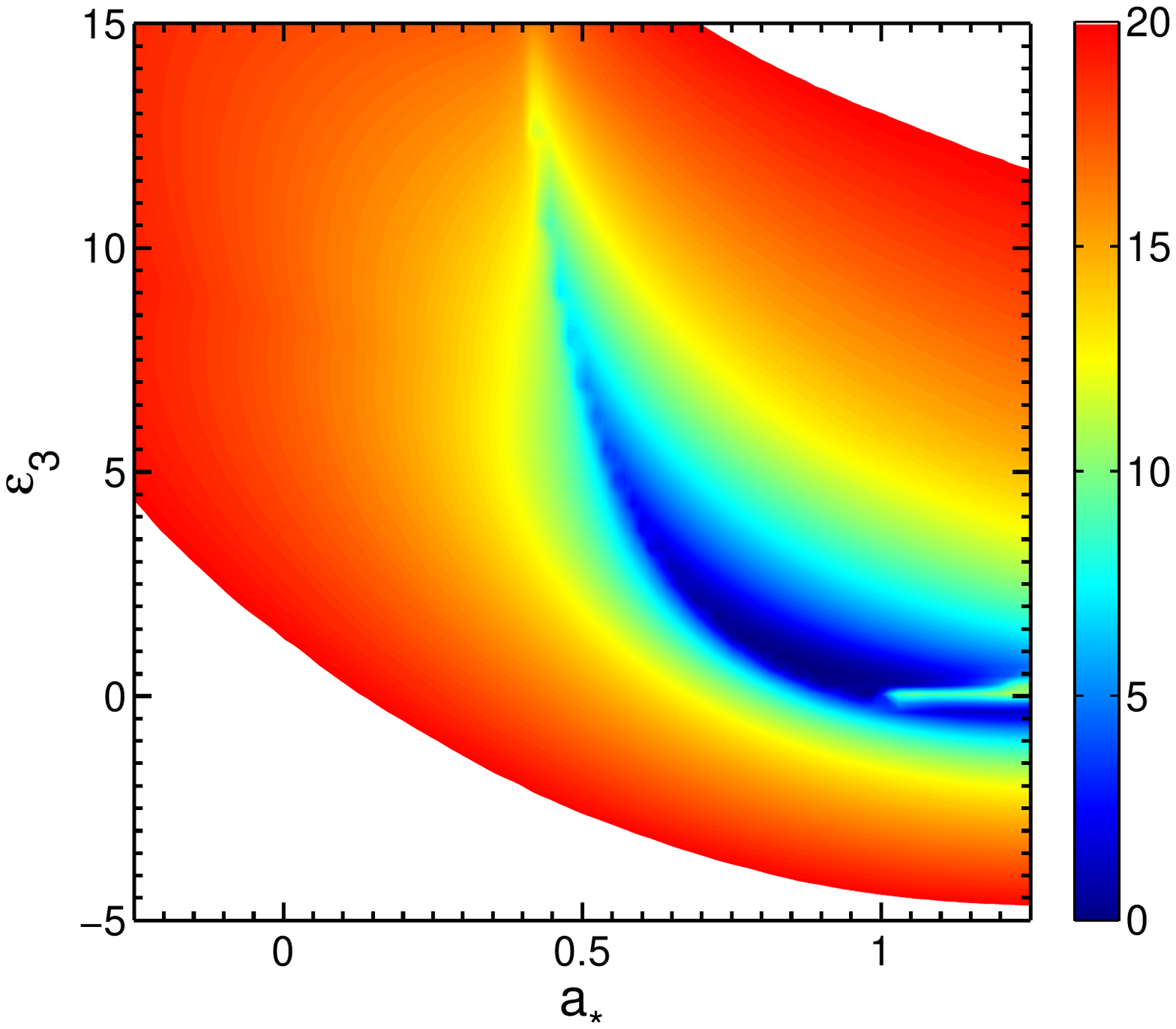}
\includegraphics[type=pdf,ext=.pdf,read=.pdf,width=8cm]{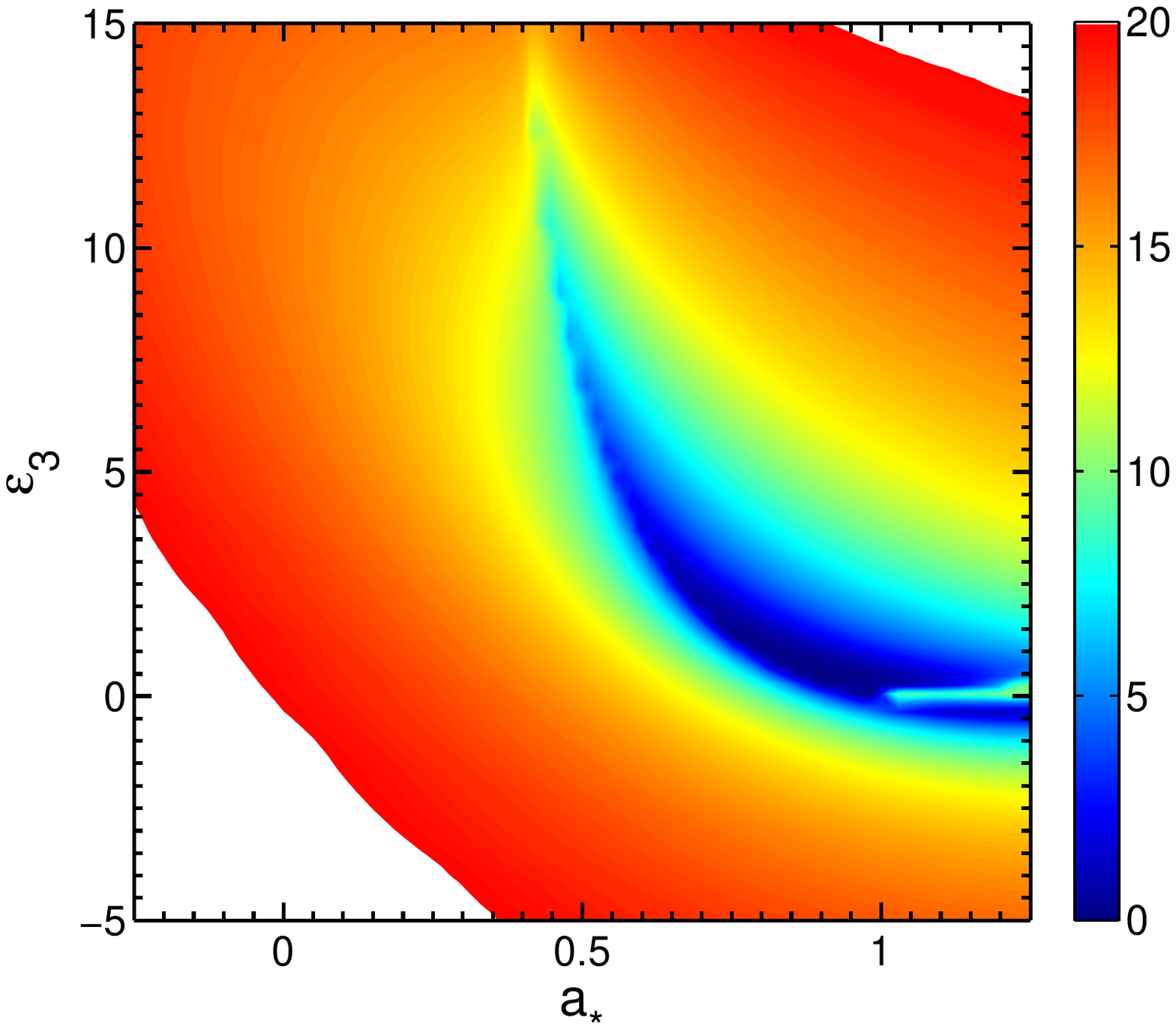} \\
\vspace{-5cm}
\includegraphics[type=pdf,ext=.pdf,read=.pdf,width=8cm]{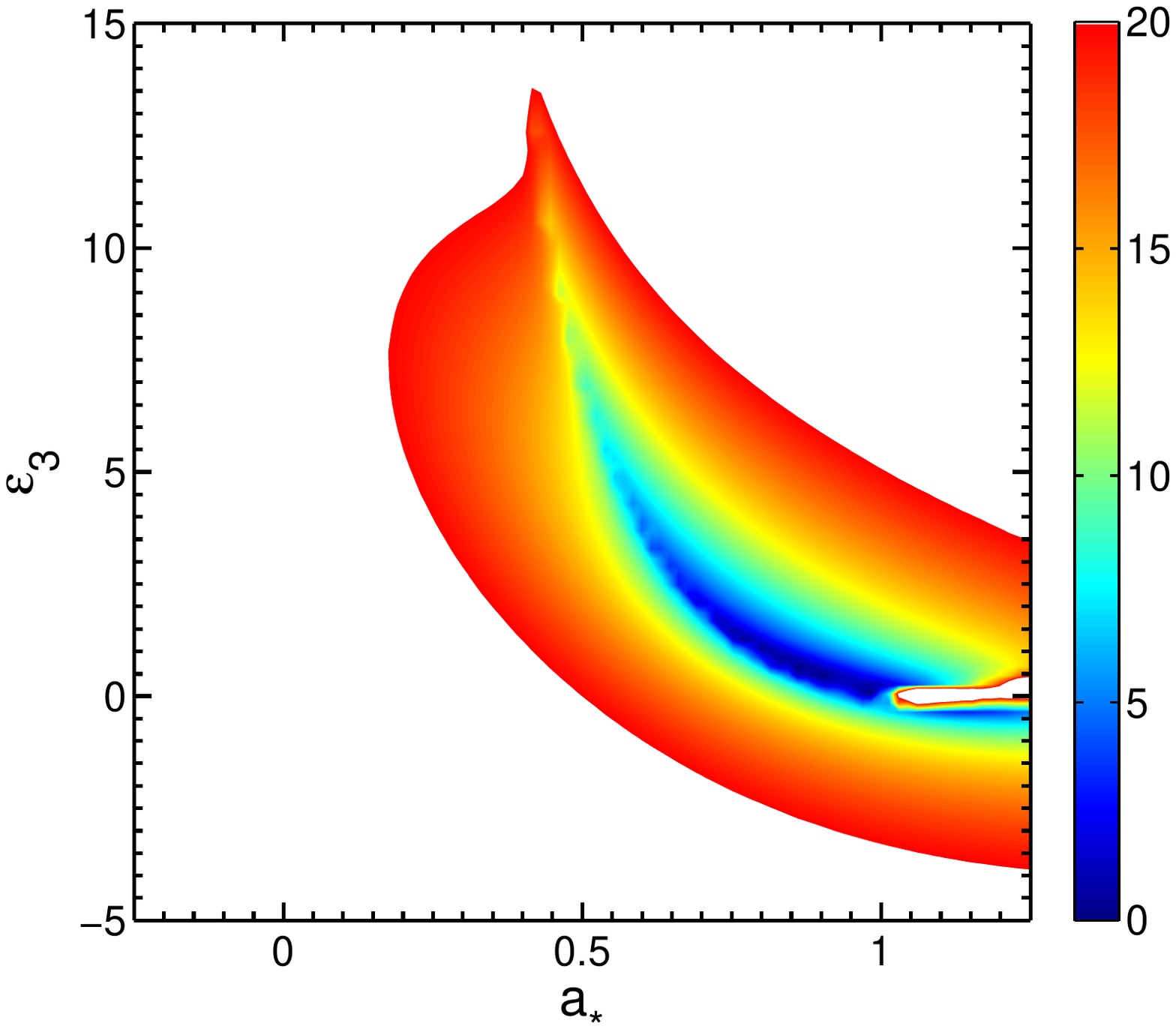}
\includegraphics[type=pdf,ext=.pdf,read=.pdf,width=8cm]{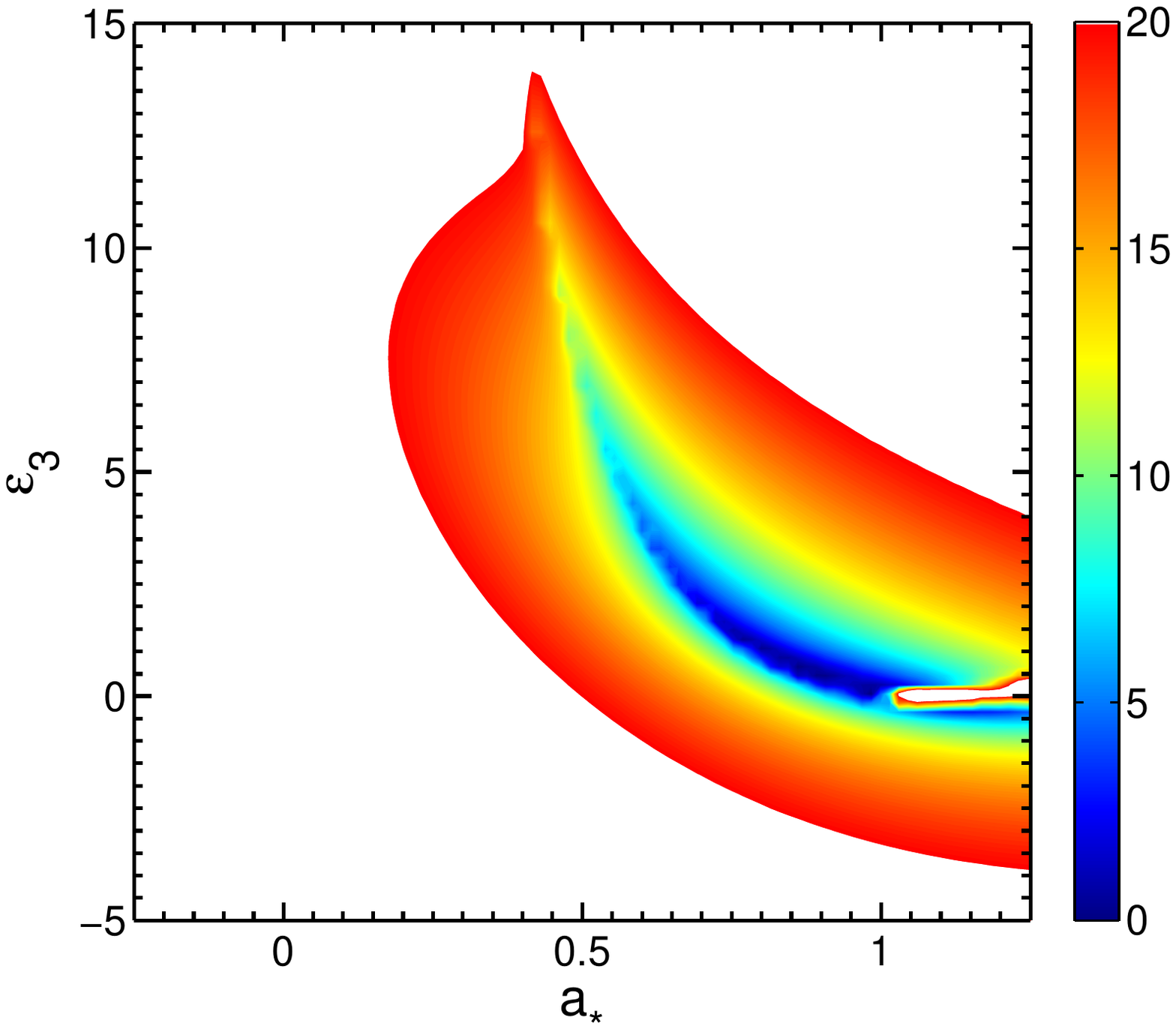} \\
\vspace{-5cm}
\includegraphics[type=pdf,ext=.pdf,read=.pdf,width=8cm]{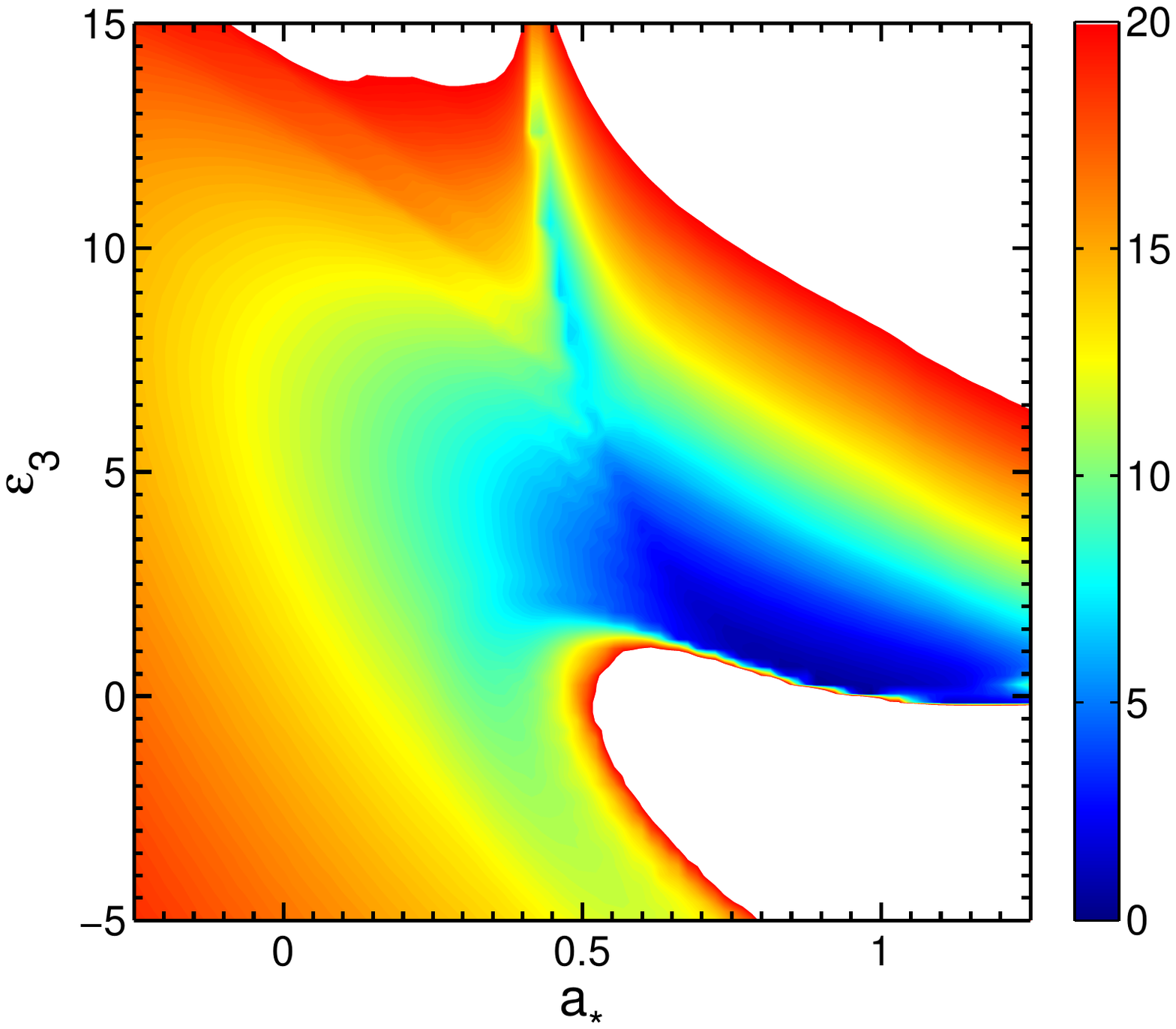}
\includegraphics[type=pdf,ext=.pdf,read=.pdf,width=8cm]{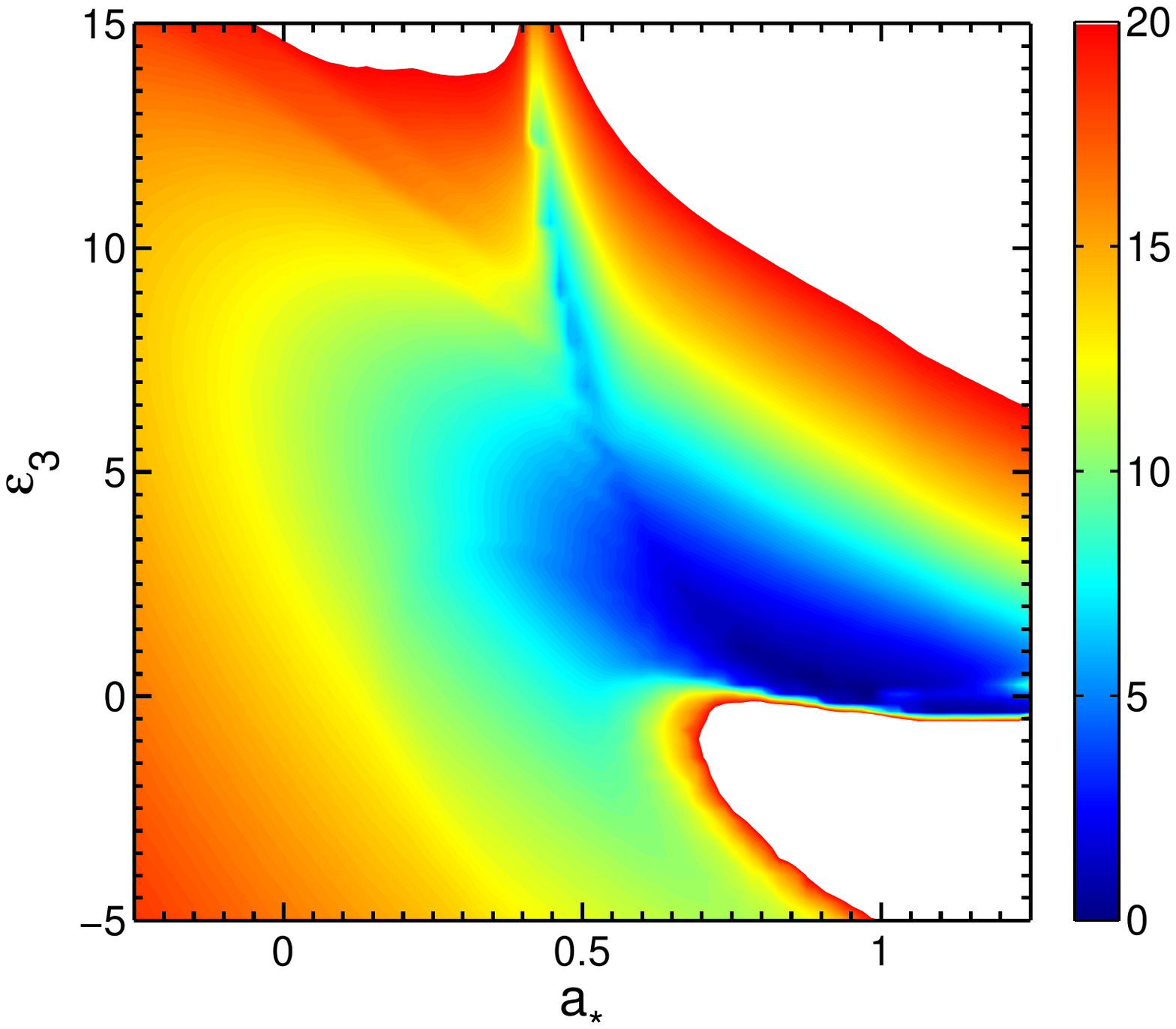}
\vspace{-2.5cm}
\caption{$\chi^2_{\rm red}$ from the comparison of the broad K$\alpha$ iron 
line produced in a Kerr space-time with spin parameter $\tilde{a}_* = 0.98$ and 
the one in a JP space-time with spin parameter $a_*$ and deformation parameter 
$\epsilon_3$. {\it Top left panel:} $\tilde{i}=i=45^\circ$. {\it Top right panel:} 
$\tilde{i}=45^\circ$ and $i$ free. {\it Central left panel:} $\tilde{i}=i=15^\circ$.
{\it Central right panel:} $\tilde{i}=15^\circ$ and $i$ free. {\it Bottom left panel:} 
$\tilde{i}=i=75^\circ$. {\it Bottom right panel:} $\tilde{i}=75^\circ$ and $i$ free.
The other parameters of the model are $\tilde{\alpha}=\alpha=-3$ and 
$\tilde{r}_{\rm out} - \tilde{r}_{\rm in}=r_{\rm out}- r_{\rm in} =100$~$M$.
See the text for details.}
\label{f-chi2-3}
\end{center}
\end{figure*}

\begin{figure*}
\begin{center}
\vspace{-2.5cm}
\includegraphics[type=pdf,ext=.pdf,read=.pdf,width=8cm]{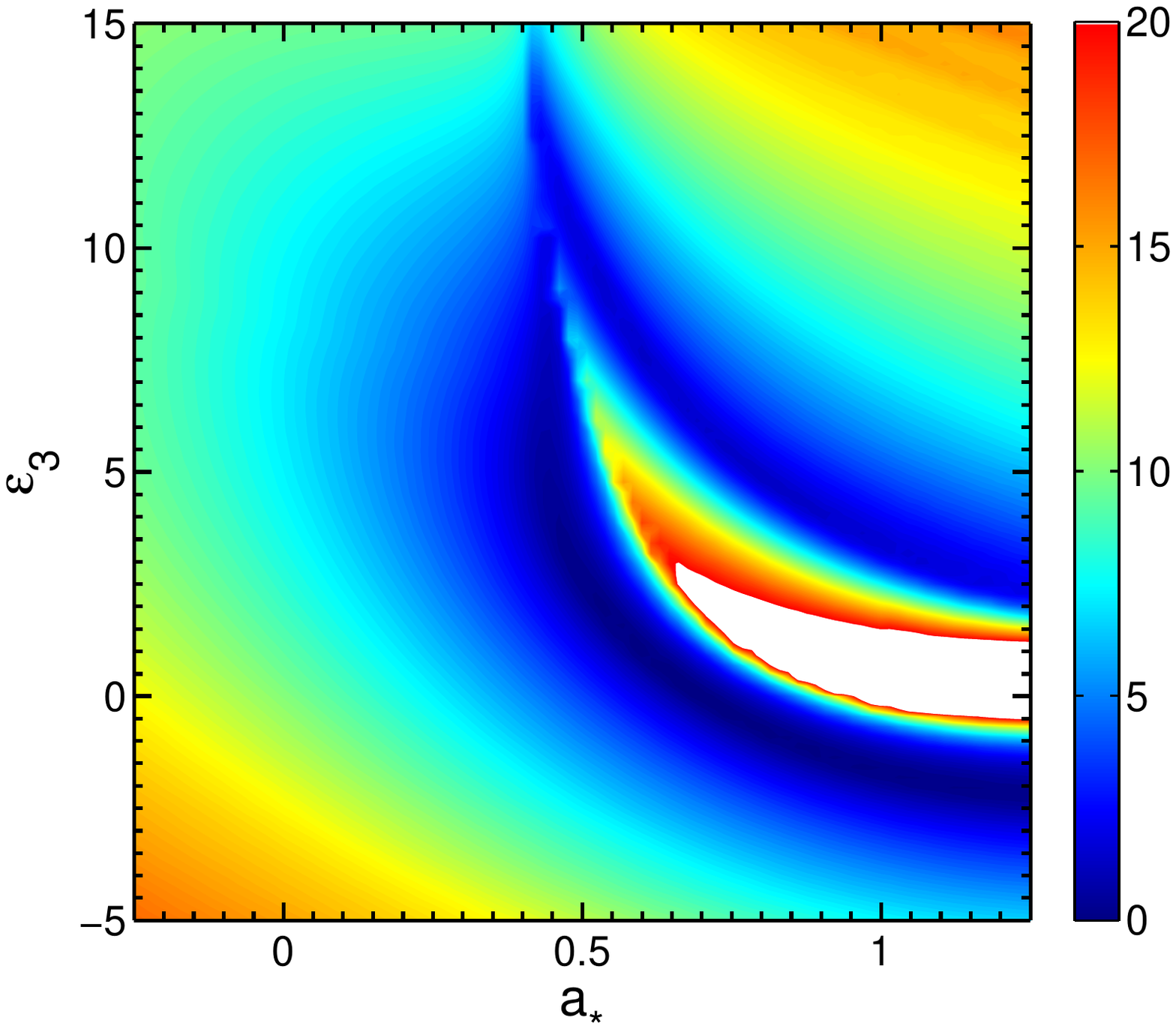}
\vspace{-2.5cm}
\caption{As in Fig.~\ref{f-chi2-1} for $\tilde{i}=i=45^\circ$, $\tilde{\alpha}=-3$, and
$\alpha$ free}
\label{f-chi2-4}
\end{center}
\end{figure*}

\section{Broad K$\alpha$ iron line in a non-Kerr background \label{s-jp}}

In order to test the Kerr-nature of astrophysical BH candidates, it is convenient
to adopt the following approach (see the first paper in~\cite{jp-em}). 
We consider a background more general 
than the Kerr metric, which includes the Kerr solution as a special case.
In the simplest scenario, we consider a metric specified by a mass $M$,
spin parameter $a_*$ and a single deformation parameter measuring possible 
deviations from the Kerr background. The idea is then to compute the
properties of the radiation emitted by the accretion disk (in our case the
shape of the K$\alpha$ iron line) in this new background and check if
observations demand a vanishing deformation parameter; that is, the 
Kerr BH hypothesis is confirmed.

The approach is clearly quite phenomenological, in the sense that we do not
aim at testing a specific and self-consistent gravity theory, but we use 
a metric as generic as possible with a deformation that, as a first approximation,
is used to quantify possible deviations from the Kerr background. The
approach makes sense because in the end we want to perform a null-experiment:
we have a deformation parameter and we want to check that it must vanish, i.e.
the compact object is a Kerr BH, regardless of the physical meaning of a 
non-zero deformation parameter. With this spirit, here I
consider the Johannsen-Psaltis (JP) metric, which can be seen as a metric
describing non-Kerr BHs in a putative alternative theory of gravity~\cite{jp-m}.
In Boyer-Lindquist coordinates, the JP metric is given by the line element
\begin{widetext}
\be\label{eq-jp}
ds^2 &=& - \left(1 - \frac{2 M r}{\Sigma}\right) (1 + h) \, dt^2
+ \frac{\Sigma (1 + h)}{\Delta + a^2 h \sin^2\theta } \, dr^2
+ \Sigma \, d\theta^2 - \frac{4 a M r \sin^2\theta}{\Sigma} (1 + h) \, dt \, d\phi + \nonumber\\
&& + \left[\sin^2\theta \left(r^2 + a^2 + \frac{2 a^2 M r \sin^2\theta}{\Sigma} \right)
+ \frac{a^2 (\Sigma + 2 M r) \sin^4\theta}{\Sigma} h \right] d\phi^2 \, ,
\ee
\end{widetext}
where $a = a_* M $, $\Sigma = r^2 + a^2 \cos^2\theta$,
$\Delta = r^2 - 2 M r + a^2$, and
\be
h = \sum_{k = 0}^{\infty} \left(\epsilon_{2k}
+ \frac{M r}{\Sigma} \epsilon_{2k+1} \right)
\left(\frac{M^2}{\Sigma}\right)^k \, .
\ee
This metric has an infinite number of deformation parameters $\epsilon_i$ and
the Kerr solution is recovered when all the deformation parameters are set to
zero. However, in order to reproduce the correct Newtonian limit, we have to
impose $\epsilon_0 = \epsilon_1 = 0$, while $\epsilon_2$ is strongly
constrained by Solar System experiments~\cite{jp-m}. In this paper, I will only 
examine the simplest cases where $\epsilon_3 \neq 0$, while all the other 
deformation parameters are set to zero.

The effects of a non-vanishing $\epsilon_3$ on the shape of the K$\alpha$ iron 
line have been discussed in Ref.~\cite{jp-fe} and are summarized in Fig.~\ref{f-jp}. 
The deformation parameter $\epsilon_3$ changes the spectrum of the line in 
a way similar to the spin parameter $a_*$, while $i$, $\alpha$, and $r_{\rm out}$ 
seem to produce different deformations. In order to be more quantitative, we
can compare the line produced in a Kerr space-time with the one
expected from a JP background. That can be done as in Ref.~\cite{cfm-b1},
by defining the reduced $\chi^2$: 
\begin{widetext}
\be\label{eq-chi2-ka}
\chi^2_{\rm red} (a_*, \epsilon_3, i, \alpha, r_{\rm out})
= \frac{\chi^2}{n} =
\frac{1}{n} \sum_{i = 1}^{n} \frac{\left[N_i^{\rm JP} 
(a_*, \epsilon_3, i, \alpha, r_{\rm out}) - N_i^{\rm Kerr}
(\tilde{a}_*, \tilde{i}, \tilde{\alpha}, \tilde{r}_{\rm out}) 
\right]^2}{\sigma^2_i} \, ,
\ee
\end{widetext} 
where the summation is performed over $n$ sampling energies $E_i$ and
$N_i^{\rm JP}$ and $N_i^{\rm Kerr}$ are the normalized photon fluxes in the 
energy bin $[E_i,E_i+\Delta E]$ respectively for the JP and Kerr metric. Here
the error $\sigma_i$ is assumed to be 15\% the normalized photon flux
$N_i^{\rm Kerr}$:
\be
\sigma_i = 0.15 \, N_i^{\rm Kerr} \, .
\ee

As a first example, we can consider the line produced around a mid-rotating Kerr
BH with $\tilde{a}_* = 0.70$. Within a simplified analysis, we can assume
$i = \tilde{i}$, $\alpha = \tilde{\alpha}$, and $r_{\rm out} - r_{\rm in} = \tilde{r}_{\rm out} 
- \tilde{r}_{\rm in}$ and make $a_*$ and $\epsilon_3$ change. This is the approach 
adopted in Ref.~\cite{jp-fe} and the motivation is that $i$, $\alpha$ and $r_{\rm out}$ 
produce different effects on the shape of the line with respect to $a_*$ and 
$\epsilon_3$ and therefore the determination of the former from the fit is relatively
independent of the measurement of the latter. Fig.~\ref{f-chi2-1} shows the
reduced $\chi^2$ for $\tilde{i}=45^\circ$ (top left panel), $\tilde{i}=15^\circ$ 
(central left panel), and $\tilde{i}=75^\circ$ (bottom left panel). The 1-$\sigma$
bound on $a_*$ and $\epsilon_3$ ($\chi^2_{\rm red} < 1$) is reported in
Tab.~\ref{tab1}. If we relax the assumption $i = \tilde{i}$, we get the plots in the
right column. The corresponding 1-$\sigma$ is clearly weaker, as shown
in Tab.~\ref{tab1}. The difficulty of constraining $\epsilon_3$ without an
independent measurement of the spin parameter $a_*$ was already pointed 
out in Ref.~\cite{jp-fe}.

If we change the value of the spin parameter of the reference spectrum, we
can find the plots in Figs.~\ref{f-chi2-2} and \ref{f-chi2-3}. Fig.~\ref{f-chi2-2}
shows the cases $\tilde{a}_* = 0.20$ (top panels) and $0.92$ (bottom panels),
for $\tilde{i}=i=45^\circ$ (left panels) and $\tilde{i}=45^\circ$ and $i$ free (right
panels). Fig.~\ref{f-chi2-3} shows instead the case $\tilde{a}_* = 0.98$
with $\tilde{i}=45^\circ$ (top panels), $\tilde{i}=15^\circ$ (central panels), and
$\tilde{i}=75^\circ$ (bottom panels), assuming either $\tilde{i}=i$ (left panels)
or with $i$ as a free parameter (right panels). The 1-$\sigma$ bounds for these
cases are reported in Tab.~\ref{tab1}.

Effects of variations of the index parameter $\alpha$ may also be important.
Fig.~\ref{f-chi2-4} shows the comparison of the iron line profile of a Kerr
BH with $\tilde{a}_* = 0.70$ with the one of a JP BH with spin $a_*$ and
deformation parameter $\epsilon_3$. In this case, I impose $i = \tilde{i}$ and 
$r_{\rm out} - r_{\rm in} = \tilde{r}_{\rm out} - \tilde{r}_{\rm in}$, while 
$\tilde{\alpha} = -3$ but $\alpha$ is free. The 1-$\sigma$ bound is only 
a little bit more stringent than the case with $\tilde{\alpha} = \alpha$ and
$i$ arbitrary. The effect of an unknown $\tilde{r}_{\rm out}$ is important for 
$\alpha = -2$ and $\tilde{r}_{\rm out} - \tilde{r}_{\rm in} \approx 10$~$M$, 
but becomes less and less relevant for lower values of the index of the
intensity profile and larger $\tilde{r}_{\rm out}$.

\begin{table*}
\begin{center}
\begin{tabular}{c c c c c c c}
\hline
\hline
Reference spectrum &  \hspace{.5cm} & Model &  \hspace{.5cm} & Kerr background &  \hspace{.5cm} & JP background \\
\hline
\hline
Kerr with $a_* = 0.70$ && $\tilde{i} = 45^\circ$ && $a_* = 0.70 \pm 0.05$ && $a_* > 0.41$, $\epsilon_3 < 5.3$ \\
\hline
 && $\tilde{i} = 45^\circ$, $i$ free && $a_* = 0.70 \pm 0.05$ && $a_* > 0.39$, $\epsilon_3 < 7.3$ \\
\hline
 && $\tilde{i} = 45^\circ$, $\alpha$ free && $a_* = 0.70 \pm 0.06$ && $a_* > 0.41$, $\epsilon_3 < 7.0$ \\
\hline
 && $\tilde{i} = 15^\circ$ && $a_* = 0.70^{+0.02}_{-0.03}$ && $a_* > 0.41$, $\epsilon_3 < 6.7$ \\
\hline
 && $\tilde{i} = 15^\circ$, $i$ free && $a_* = 0.70 \pm 0.03$ && $a_* > 0.39$, $\epsilon_3 < 7.8$ \\
\hline
 && $\tilde{i} = 75^\circ$ && $a_* = 0.70 \pm 0.09$ && $a_* > 0.53$, $\epsilon_3 < 1.8$ \\
\hline
 && $\tilde{i} = 75^\circ$, $i$ free && $a_* = 0.70 \pm 0.09$ && $a_* > 0.48$, $\epsilon_3 < 2.6$ \\
\hline
 && $\tilde{i} = 45^\circ$, CFM && $a_* = 0.70^{+0.03}_{-0.05}$ && no bound \\
\hline
 && $\tilde{i} = 45^\circ$, CFM+K$\alpha$ && $a_* = 0.70 \pm 0.03$ && $0.45 < a_* < 1.14$, $-2.3< \epsilon_3 < 2.8$ \\
\hline
\hline
Kerr with $a_* = 0.20$ && $\tilde{i} = 45^\circ$ && $a_* = 0.20^{+0.05}_{-0.10}$ && $a_* > -0.12$, $\epsilon_3 < 7.4$ \\
\hline
 && $\tilde{i} = 45^\circ$, $i$ free && $a_* = 0.20^{+0.05}_{-0.10}$ && $a_* > -0.18$, $\epsilon_3 < 11.6$ \\
\hline
\hline
Kerr with $a_* = 0.92$ && $\tilde{i} = 45^\circ$ && $a_* = 0.92^{+0.03}_{-0.04}$ && $a_* > 0.53$, $\epsilon_3 < 5.6$ \\
\hline
 && $\tilde{i} = 45^\circ$, $i$ free && $a_* = 0.92^{+0.03}_{-0.04}$ && $a_* > 0.49$, $\epsilon_3 < 7.3$ \\
\hline
\hline
Kerr with $a_* = 0.98$ && $\tilde{i} = 45^\circ$ && $a_* = 0.98^{+0.01}_{-0.03}$ && $a_* > 0.65$, $\epsilon_3 < 2.6$ \\
\hline
 && $\tilde{i} = 45^\circ$, $i$ free && $a_* = 0.98^{+0.01}_{-0.03}$ && $a_* > 0.63$, $\epsilon_3 < 3.1$ \\
\hline
 && $\tilde{i} = 15^\circ$ && $a_* = 0.98 \pm 0.01$ && $0.71 < a_* < 1.10$, $-0.2 < \epsilon_3 < 1.8$ \\
\hline
 && $\tilde{i} = 15^\circ$, $i$ free && $a_* = 0.98 \pm 0.01$ && $0.71 < a_* < 1.15$, $-0.3 < \epsilon_3 < 1.8$ \\
\hline
 && $\tilde{i} = 75^\circ$ && $a_* = 0.98 \pm 0.01$ && $a_* > 0.77$, $\epsilon_3 < 1.2$ \\
\hline
 && $\tilde{i} = 75^\circ$, $i$ free && $a_* = 0.98^{+0.01}_{-0.07}$ && $a_* > 0.74$, $\epsilon_3 < 1.5$ \\
\hline
 && $\tilde{i} = 45^\circ$, CFM && $a_* = 0.98^{+0.01}_{-0.02}$ && $a_* > 0.44$, $\epsilon_3 < 12.4$ \\
\hline
 && $\tilde{i} = 45^\circ$, CFM+K$\alpha$ && $a_* = 0.98 \pm 0.01$ && $0.70 < a_* < 1.10$, $-0.4< \epsilon_3 < 2.2$ \\
\hline
\hline
\end{tabular}
\end{center}
\vspace{-0.2cm}
\caption{1-$\sigma$ bounds on the spin parameter $a_*$ (in the Kerr background) and on
the spin parameter-deformation parameter plane (in the JP background) for the cases
shown in Figs.~\ref{f-chi2-1}-\ref{f-chi2-6}. Unless stated otherwise, $\tilde{i}=i$, 
$\tilde{\alpha}=\alpha=-3$, and $\tilde{r}_{\rm out} - \tilde{r}_{\rm in}=r_{\rm out}- 
r_{\rm in} =100$~$M$. CFM stands for continuum-fitting method, while CFM+K$\alpha$
is the case in which the measurements from the continuum-fitting method and the K$\alpha$
iron line are combined. In this paper, I studied only the region $-0.25 < a_* < 1.25$ and 
$-5 < \epsilon_3 < 15$ of the spin-deformation parameter plane; the expression ``no bound''
in the table means that at 1-$\sigma$ it is not possible to exclude any value of the spin 
in the range $-0.25 < a_* < 1.25$, as well as any value of the deformation parameter in the range 
$-5 < \epsilon_3 < 15$.}
\label{tab1}
\end{table*}

\begin{table*}
\begin{center}
\begin{tabular}{c c c c c c c}
\hline
\hline
Reference spectrum &  \hspace{.5cm} & Model &  \hspace{.5cm} & $\epsilon_3 = 7.0$ JP background &  \hspace{.5cm} & JP background \\
\hline
\hline
JP with $a_* = 0.20$, $\epsilon_3 = 7.0$ && $\tilde{i} = 45^\circ$ && $a_* = 0.20 \pm 0.05$ && $0.15 < a_* < 0.25$, $6.3 < \epsilon_3 < 8.8$ \\
\hline
 && $\tilde{i} = 45^\circ$, $i$ free && $a_* = 0.20 \pm 0.05$ && $0.15 < a_* < 0.86$, $\epsilon_3 < 11.2$ \\
\hline
 && $\tilde{i} = 45^\circ$, CFM && $a_* = 0.20^{+0.03}_{-0.04}$ && no bound \\
\hline
 && $\tilde{i} = 45^\circ$, CFM+K$\alpha$ && $a_* = 0.20 \pm 0.03$ && $0.17 < a_* < 0.24$, $6.3< \epsilon_3 < 7.7$ \\
\hline
\hline
\end{tabular}
\end{center}
\vspace{-0.2cm}
\caption{As in Tab.~\ref{tab1} for the cases shown in Fig.~\ref{f-nk}, in which the reference
metric has $a_* = 0.20$ and $\epsilon_3 = 7.0$.}
\label{tab2}
\end{table*}

\begin{figure*}
\begin{center}
\vspace{-2.5cm}
\includegraphics[type=pdf,ext=.pdf,read=.pdf,width=8cm]{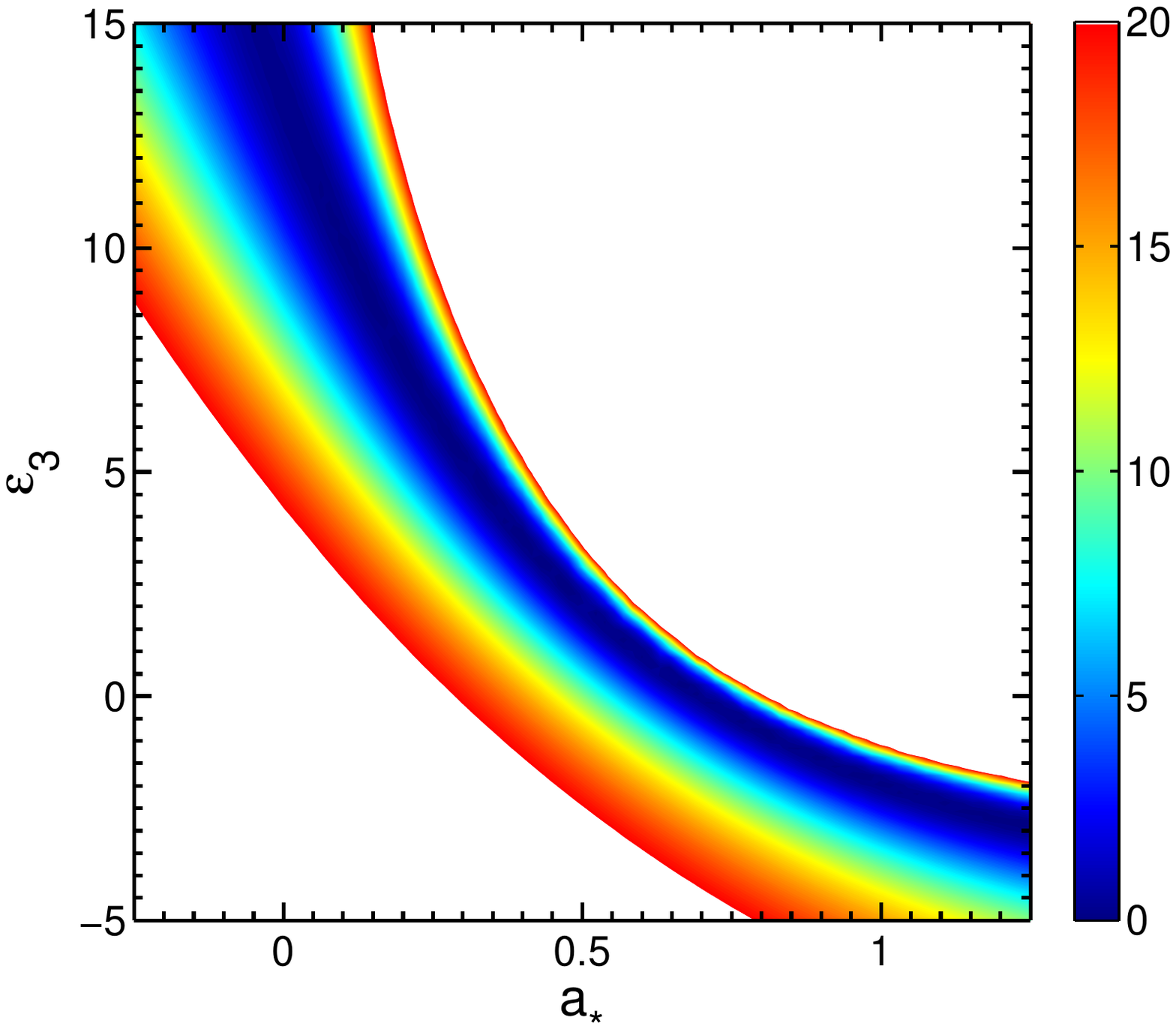}
\includegraphics[type=pdf,ext=.pdf,read=.pdf,width=8cm]{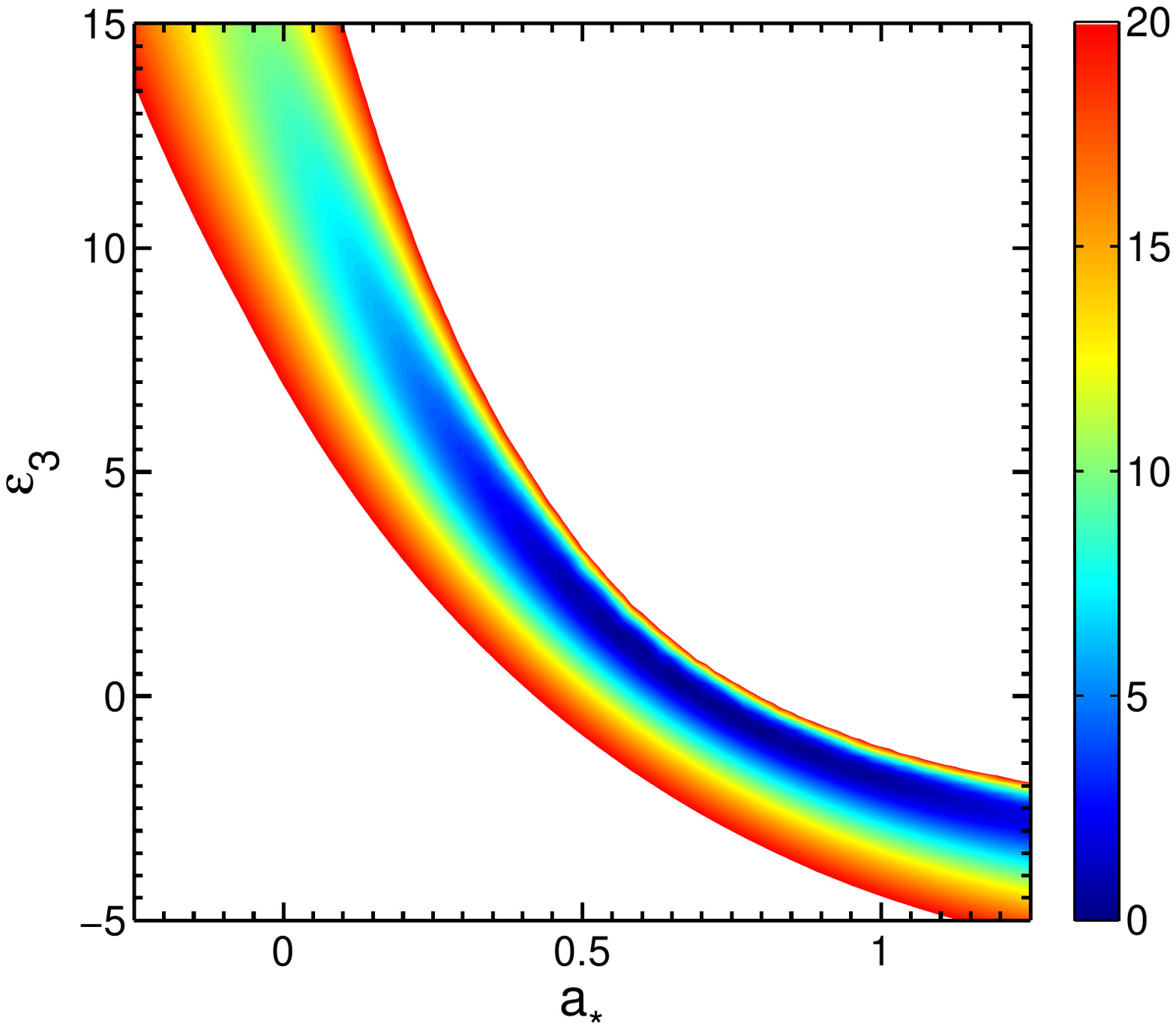}
\vspace{-2.5cm}
\caption{{\it Left panel:} $\chi^2_{\rm red}$ from the comparison of the thermal 
spectrum of a thin accretion disk around a Kerr BH with spin parameter 
$\tilde{a}_* = 0.70$ and the one in a JP space-time with spin parameter $a_*$ and 
deformation parameter $\epsilon_3$. {\it Right panel:} $\chi^2_{\rm red,\, tot}$ 
from the combination of the analysis of the broad K$\alpha$ iron line and the 
continuum-fitting method. The parameters of the model are: mass $M = 10$~$M_\odot$, 
mass accretion rate $\dot{M} = 2 \cdot 10^{18}$~g~s$^{-1}$, distance $D = 10$~kpc, 
and viewing angle $i = 45^\circ$. See the text for details.}
\label{f-chi2-5}
\end{center}
\end{figure*}

\begin{figure*}
\begin{center}
\vspace{-2.5cm}
\includegraphics[type=pdf,ext=.pdf,read=.pdf,width=8cm]{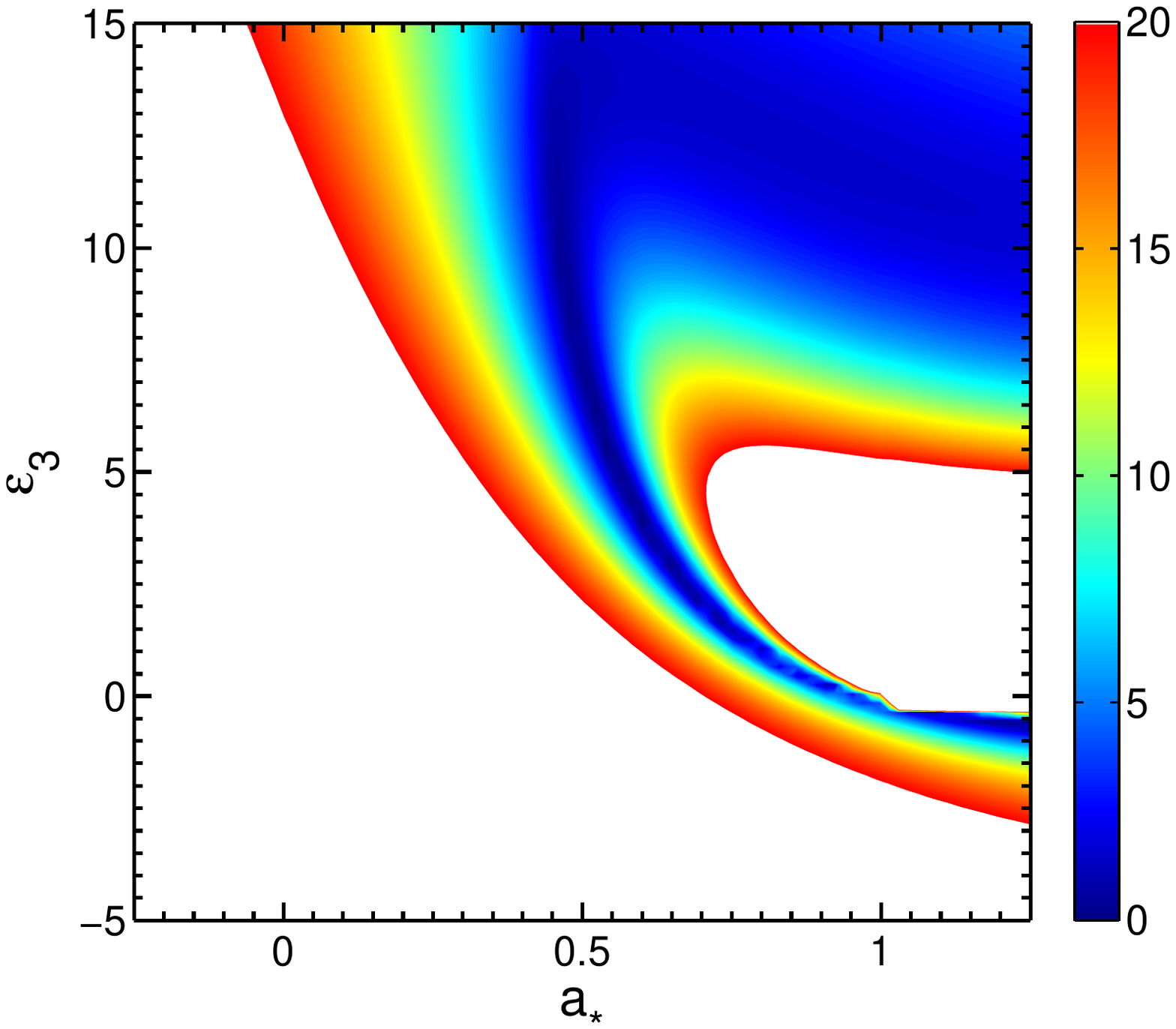}
\includegraphics[type=pdf,ext=.pdf,read=.pdf,width=8cm]{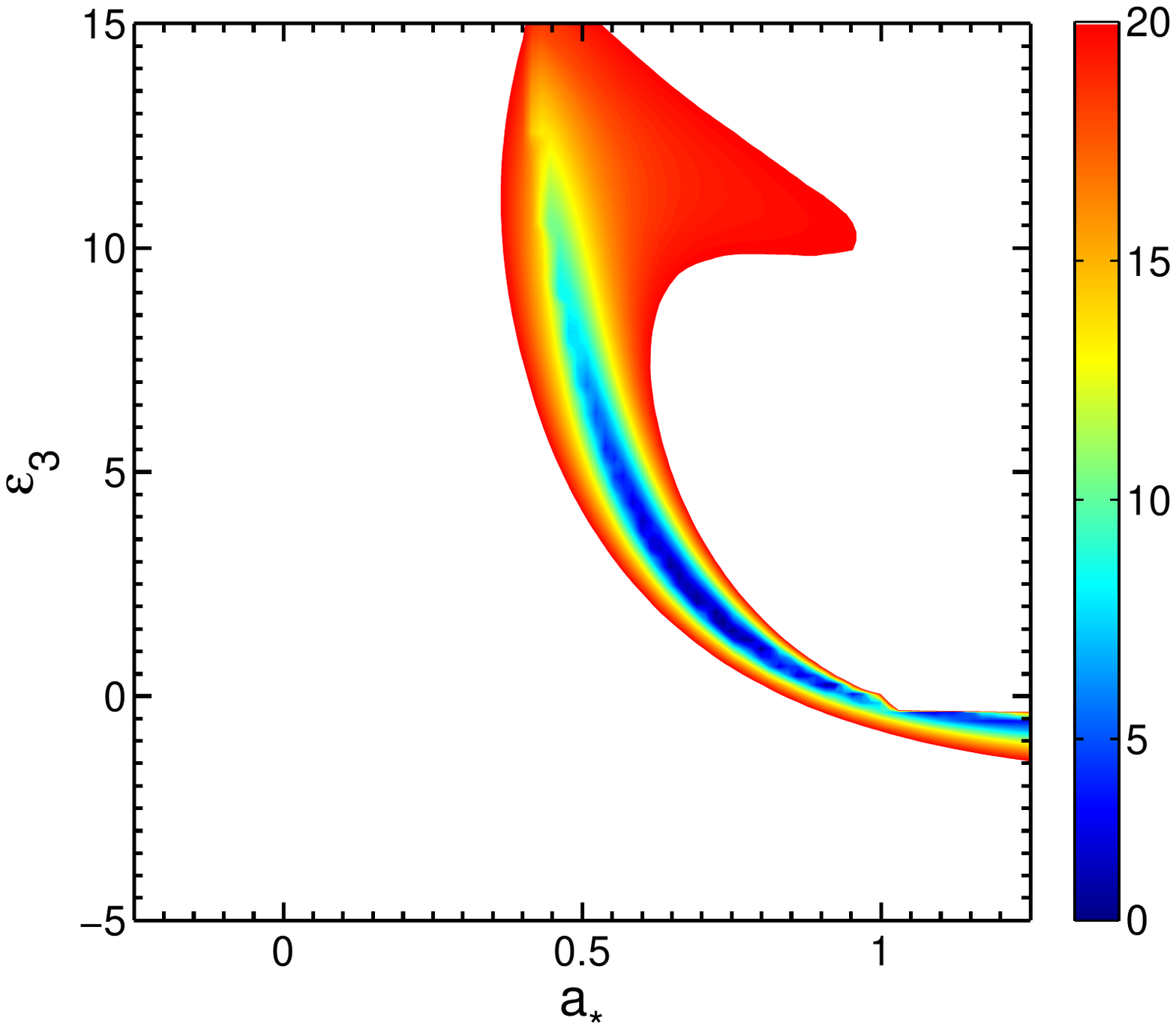}
\vspace{-2.5cm}
\caption{{\it Left panel:} $\chi^2_{\rm red}$ from the comparison of the thermal 
spectrum of a thin accretion disk around a Kerr BH with spin parameter 
$\tilde{a}_* = 0.98$ and the one in a JP space-time with spin parameter $a_*$ and 
deformation parameter $\epsilon_3$. {\it Right panel:} $\chi^2_{\rm red,\, tot}$ 
from the combination of the analysis of the broad K$\alpha$ iron line and the 
continuum-fitting method. The parameters of the model are: mass $M = 10$~$M_\odot$, 
mass accretion rate $\dot{M} = 2 \cdot 10^{18}$~g~s$^{-1}$, distance $D = 10$~kpc, 
and viewing angle $i = 45^\circ$. See the text for details.}
\label{f-chi2-6}
\end{center}
\end{figure*}

\begin{figure*}
\begin{center}
\vspace{-2.5cm}
\includegraphics[type=pdf,ext=.pdf,read=.pdf,width=8cm]{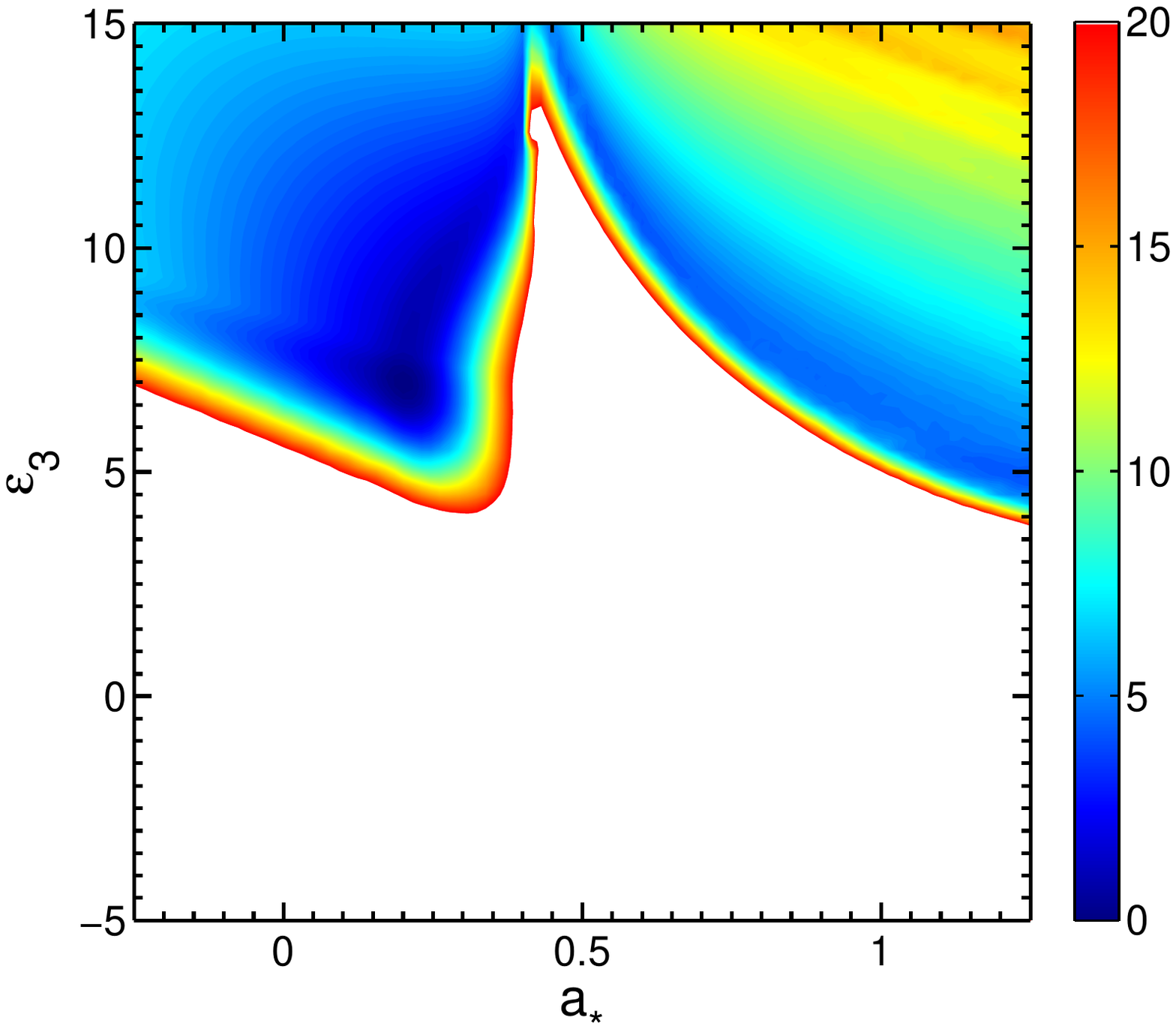}
\includegraphics[type=pdf,ext=.pdf,read=.pdf,width=8cm]{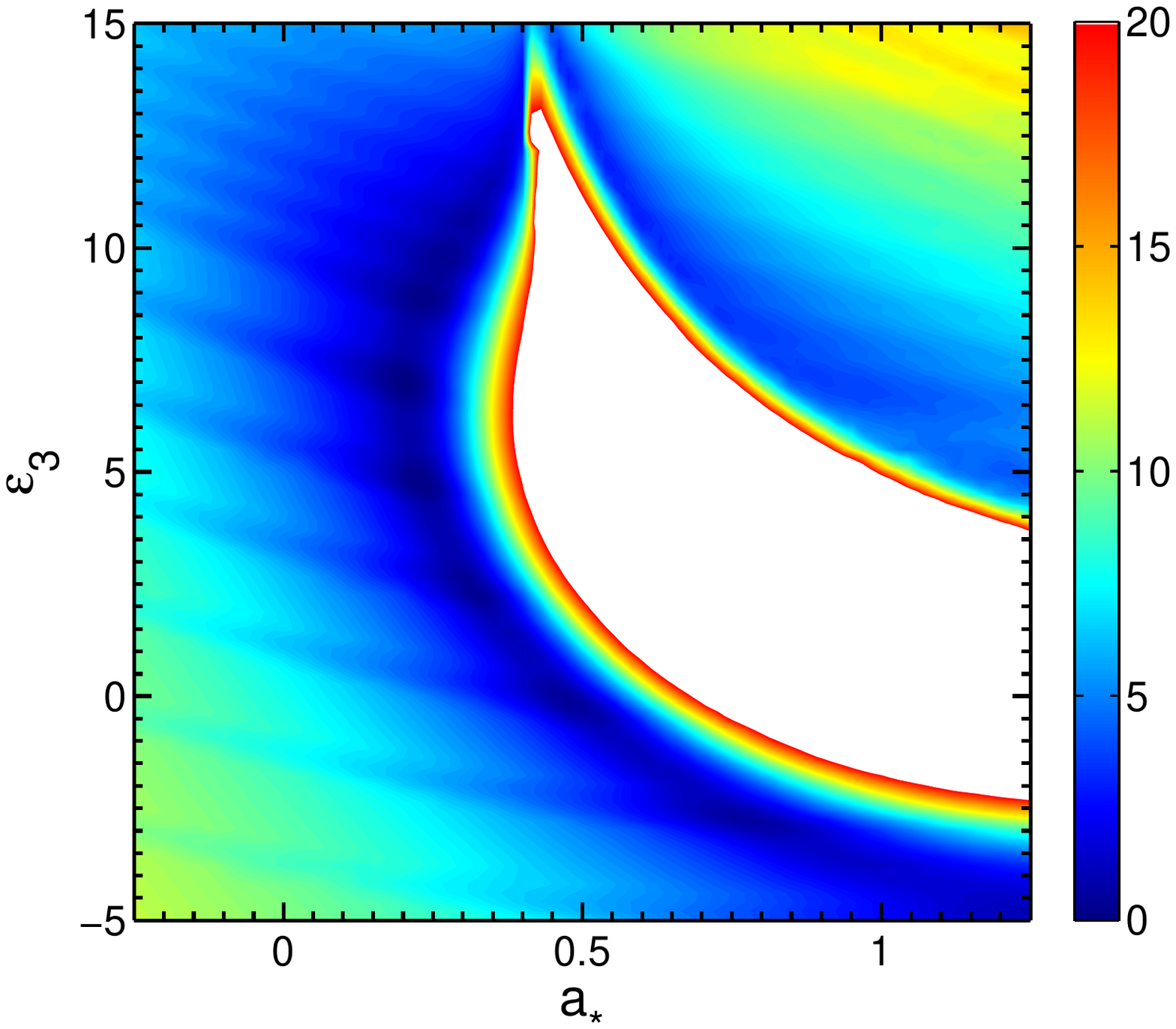} \\
\vspace{-5cm}
\includegraphics[type=pdf,ext=.pdf,read=.pdf,width=8cm]{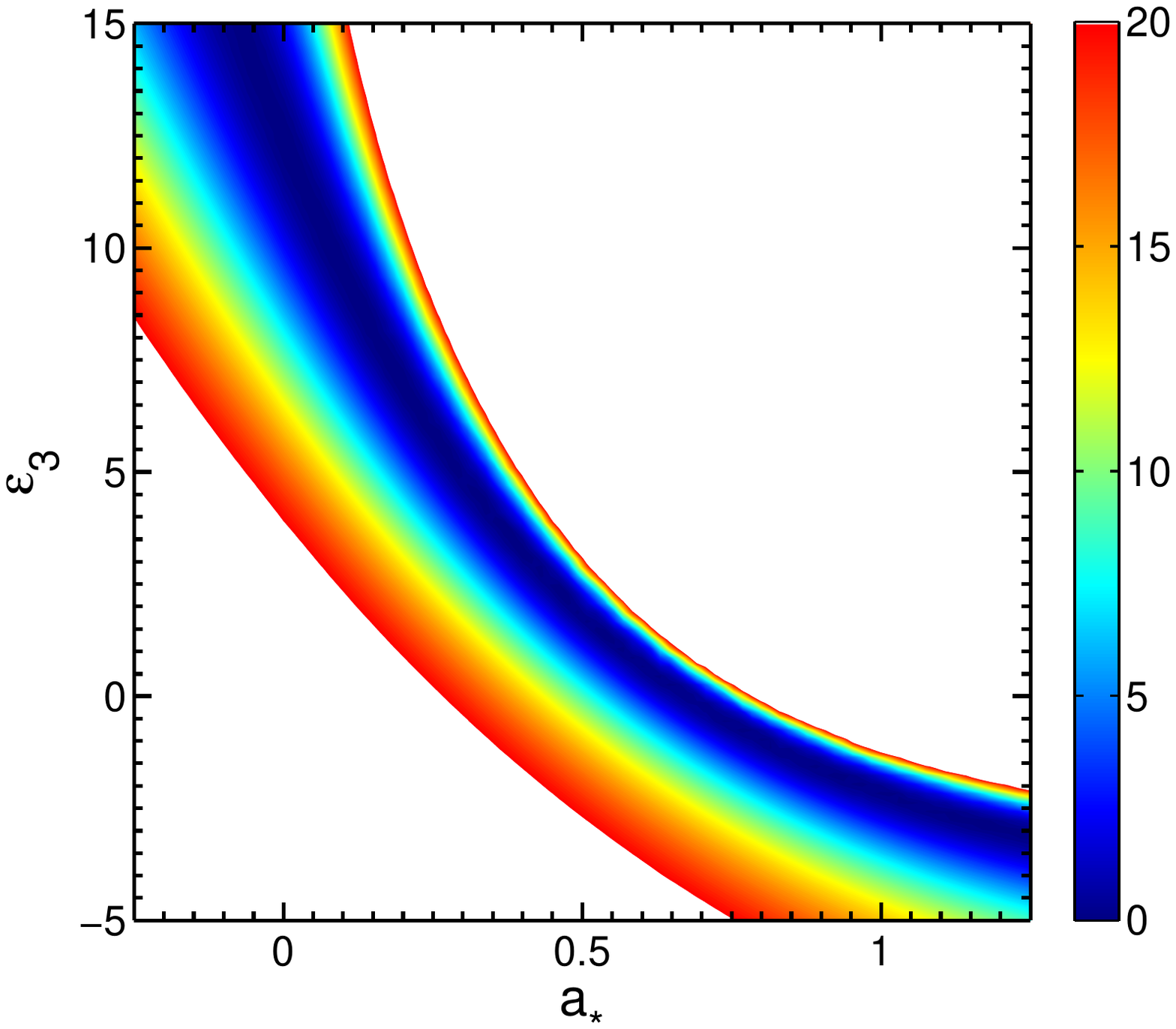}
\includegraphics[type=pdf,ext=.pdf,read=.pdf,width=8cm]{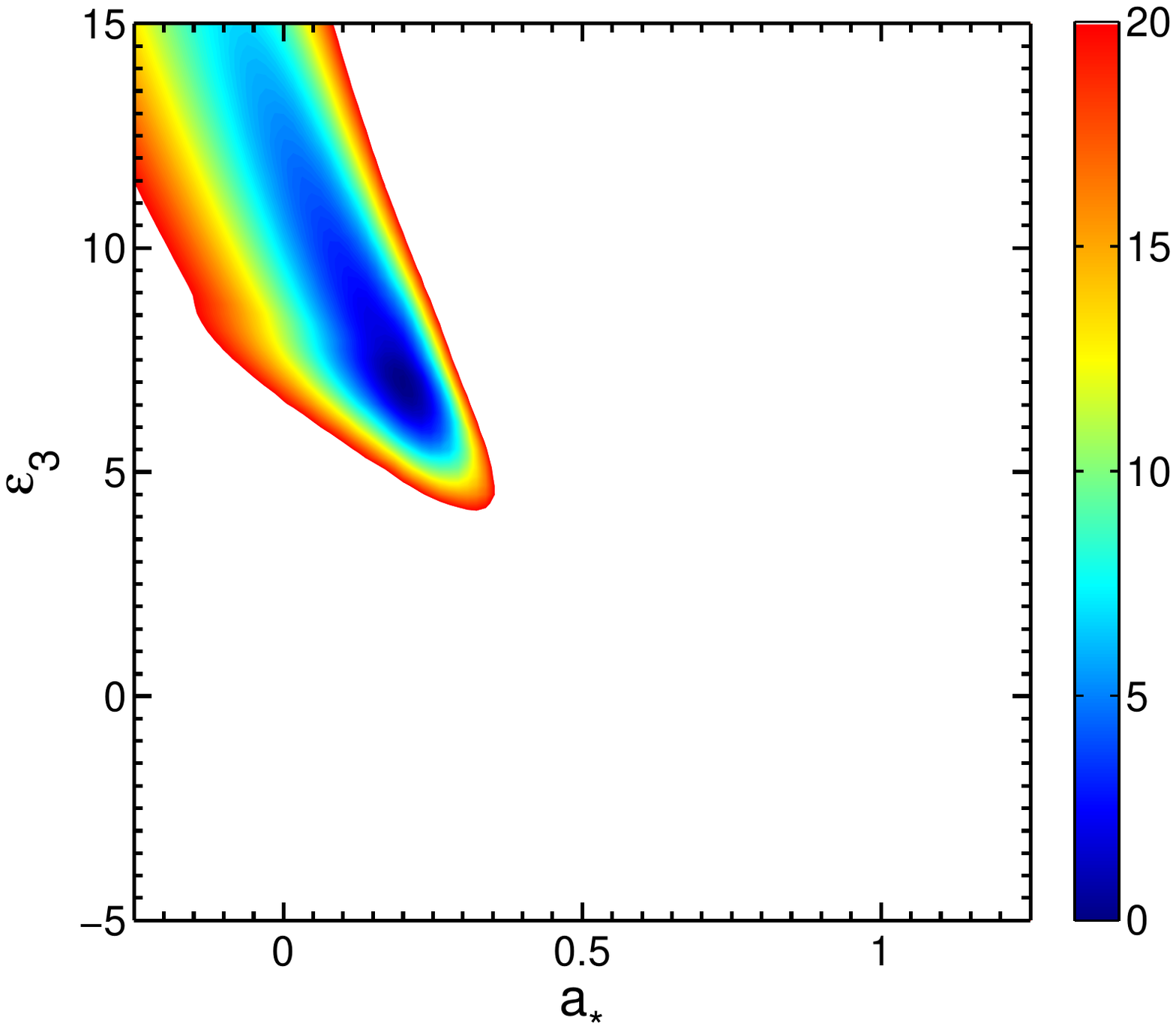}
\vspace{-2.5cm}
\caption{{\it Top left panel:} $\chi^2_{\rm red}$ from the comparison of the broad 
K$\alpha$ iron line produced in a JP space-time with spin parameter 
$\tilde{a}_* = 0.20$ and $\tilde{\epsilon}_3 = 7.0$ and the one in a JP space-time 
with spin parameter $a_*$ and deformation parameter $\epsilon_3$. 
$\tilde{i}=i=45^\circ$. {\it Top right panel:} as in the left panel, for $\tilde{i}=45^\circ$
and $i$ free. {\it Bottom left panel:} $\chi^2_{\rm red}$ from the comparison of the 
thermal spectrum of a thin accretion disk around a JP BH with spin parameter 
$\tilde{a}_* = 0.20$ and $\tilde{\epsilon}_3 = 7.0$ and the one in a JP space-time 
with spin parameter $a_*$ and deformation parameter $\epsilon_3$.  The parameters 
of the model are: mass $M = 10$~$M_\odot$, mass accretion rate $\dot{M} = 
2 \cdot 10^{18}$~g~s$^{-1}$, distance $D = 10$~kpc, and viewing angle $i = 45^\circ$.
{\it Bottom right panel:} $\chi^2_{\rm red,\, tot}$ from the combination of the 
analysis of the broad K$\alpha$ iron line and the continuum-fitting method. See 
the text for details.}
\label{f-nk}
\end{center}
\end{figure*}

\section{Discussion \label{s-e}}

\subsection{Constraints on the spin parameter-deformation parameter plane}

In the Kerr background, the exact value of the spin parameter sets the
inner edge of the accretion disk. The radiation emitted from the region
closer to the BH candidate is more gravitationally redshifted and for this reason
$a_*$ can be estimated by studying the low-energy tail of the broad 
K$\alpha$ iron line. On the other hand, the Keplerian 
velocity profile depends very weakly on the spin parameter. The 
high-energy peak of the line is produced by the Doppler boosting 
from radiation emitted at relatively large radii; indeed, if the emissivity
at large radii is strongly suppressed the high-energy peak disappears
(e.g. for $\alpha \lesssim -4$, see the left bottom panel of Fig.~\ref{f-kerr}). 
The position of the high-energy peak is thus determined 
by the inclination angle of the disk with respect to the observer's line of sight.
In the Kerr space-time, the spin parameter and the inclination angle can 
be determined by two different parts of the relativistically broadened
line and the two measurements are not correlated.

If we consider the possibility of a non-vanishing deformation parameter,
the picture is different. In the case of the JP background discussed in the
previous section, the effect of a small $\epsilon_3$ is very similar to
a variation of the spin parameter $a_*$; that is, $\epsilon_3$ alters the
low-energy tail of the line and it is not possible to distinguish a line
emitted around a Kerr BH with the one coming from a non-Kerr compact
object with a different spin parameter. However, for larger deviations
from the Kerr geometry, the variation of the Keplerian velocity profile
also becomes relevant. That has two effects. First, the distortion produced
by $a_*$ and $\epsilon_3$ are not equivalent any more and it is possible to 
get a bound on possible deviations from Kerr independently of the
value of the spin parameter. For instance, in the case of the
thermal spectrum of the disk, interesting bounds may not be possible 
(see next subsection). Second, the determination
of $a_*$ and $\epsilon_3$ is not really independent of the measurement
of the inclination angle $i$: for this reason,
the fact the inclination angle has also to be determined by the fit increases
the allowed region in the plane spin parameter-deformation parameter.

\subsection{Comparison with the continuum-fitting method}

It is interesting to compare the information on the space-time geometry 
present in the shape of the broad K$\alpha$ iron line with the one in
the thermal spectrum of a thin accretion disk. The extension of the 
continuum-fitting method to non-Kerr space-times was 
discussed in Refs.~\cite{cfm-b1,cb-apj}, where it was shown that this technique 
actually measures the radiative efficiency of the Novikov-Thorne model, 
$\eta = 1 - E_{\rm ISCO}$. In other words, the approach cannot really distinguish 
the spectrum of a thin disk around a Kerr BH with spin parameter $a_*$ and 
radiative efficiency $\eta$ from the one in a non-Kerr space-time with 
completely different value of the spin parameter but same radiative efficiency.
This simple rule was exploited in~\cite{cfm-b2} to quickly get the allowed region on
the spin parameter-deformation parameter plane of the stellar-mass BH 
candidates whose spin has been estimated under the assumption 
of the Kerr metric.

The thermal spectrum of a thin disk in the Kerr space-time depends on five 
parameters (mass of the BH candidate $M$, spin parameter $a_*$, mass 
accretion rate of the BH candidate $\dot{M}$, distance of the binary system 
$D$, and inclination angle of the disk $i$). $M$, $D$, and $i$ should be 
deduced from independent measurements, while $a_*$ and $\dot{M}$ are 
inferred by fitting the disk's spectrum. However, $a_*$ and $\dot{M}$ are not 
correlated, as the former changes the position of the peak, while the latter changes 
the intensity of the low energy region of the spectrum. We can then define 
$\chi^2_{\rm red}$ with the same spirit of Section~\ref{s-jp} and 
Ref.~\cite{cfm-b1}
\be\label{eq-chi2-cfm}
\chi^2_{\rm red} (a_*, \epsilon_3) =
\frac{1}{n} \sum_{i = 1}^{n} \frac{\left[N_i^{\rm JP} 
(a_*, \epsilon_3) - N_i^{\rm Kerr} (\tilde{a}_*) 
\right]^2}{\sigma^2_i} \, , \nonumber\\
\ee 
and compare the disk's spectrum of a Kerr BH with the one of a JP space-time.
The uncertainty $\sigma_i$ is still assumed to be 15\% the photon flux of the 
reference spectrum; i.e., $\sigma_i = 0.15 \, N_i^{\rm Kerr} (\tilde{a}_*)$.
Here the spectra have been computed with the code described in Ref.~\cite{cb-apj}, 
assuming a mass accretion rate $\dot{M} = 2 \cdot 10^{18}$~g~s$^{-1}$ and 
a viewing angle $i = 45^\circ$. Unlike the analysis of the broad iron line
profile, in the continuum-fitting method the inclination angle $i$ must be obtained
from independent measurements and is an input parameter. The resulting
$\chi^2_{\rm red}$ are shown in the left panels of Figs.~\ref{f-chi2-5} and \ref{f-chi2-6},
respectively for a spin parameter $\tilde{a}_* = 0.70$ and 0.98. The 1-$\sigma$
constraints are reported in Tab.~\ref{tab1}. Especially
for $\tilde{a}_* = 0.70$, deviations from the Kerr background can be very large 
if we increase the difference of the value of $a_*$ with respect to the Kerr case.
At the same time, for small values of the deformation parameter $\epsilon_3$,
the degeneracy between $a_*$ and $\epsilon_3$ in the continuum-fitting 
method and in the analysis of the broad K$\alpha$ iron line is very similar.

If we have a BH candidate for which we can get good data of the thermal 
spectrum of its accretion disk in the high-soft state and of the broad
K$\alpha$ iron line, we can combine the two measurements. We can see 
the constraint we can obtain from the combinations of the two techniques
by introducing $\chi^2_{\rm red, \, tot}$, defined as
\be
\chi^2_{\rm red,\, tot}  = 
\chi^2_{\rm red,\, K\alpha} + \chi^2_{\rm red,\, cfm} \, ,
\ee
where $\chi^2_{\rm red,\, K\alpha}$ and $\chi^2_{\rm red,\, cfm}$ are, respectively, 
the reduced $\chi^2$ from the analysis of the K$\alpha$ iron line, defined in
Eq.~(\ref{eq-chi2-ka}), and of the thermal spectrum of a thin accretion disk, 
defined in Eq.~(\ref{eq-chi2-cfm}). If we consider the case in which 
$\tilde{i} =i= 45^\circ$, $\tilde{\alpha} = \alpha = -3$, and $r_{\rm out} - r_{\rm in} = 
\tilde{r}_{\rm out} - \tilde{r}_{\rm in} = 100$~$M$, the $\chi^2_{\rm red,\, tot}$ is 
the one shown in the right panel of Fig.~\ref{f-chi2-5} for $\tilde{a}_* = 0.70$
and in the right panel of Fig.~\ref{f-chi2-6} for $\tilde{a}_* = 0.98$. The 1-$\sigma$
constraints are reported in Tab.~\ref{tab1}.

Lastly, we can consider the possibility that astrophysical BH candidates are not 
Kerr BHs and therefore that the reference spectrum is generated in a space-time
with non-vanishing deformation parameter. $\chi^2_{\rm red}$ is now given by
Eqs.~(\ref{eq-chi2-ka}) and (\ref{eq-chi2-cfm}) with $N_i^{\rm Kerr}$ replaced by
$N_i^{\rm JP}(\tilde{a_*}, \tilde{\epsilon_3}, ...)$. Fig.~\ref{f-nk} shows the results
for the case with $\tilde{a}_* = 0.20$ and $\tilde{\epsilon}_3 = 7.0$. In the top
left panel, there is the $\chi^2_{\rm red}$ from the K$\alpha$ iron line, assuming
$\tilde{i}=i=45^\circ$, $\tilde{\alpha} = \alpha = -3$, and $r_{\rm out} - r_{\rm in} = 
\tilde{r}_{\rm out} - \tilde{r}_{\rm in} = 100$~$M$. Here, if we try to fit the line profile
with the one of a Kerr space-time, we should get a bad fit, at least if all the
astrophysical processes are properly taken into account and one of the latter is
not used to make the fit good. However, as $i$ is also a fit parameter, the
assumption $\tilde{i}=i$ is not justified. If we consider $i$ as a free parameter,
the picture changes dramatically. The reduced $\chi^2$ from the iron line is now 
shown in the top right panel of Fig.~\ref{f-nk} and it is perfectly consistent with a 
Kerr BH with spin $a_* \sim 0.5$. If we repeat the analysis of the disk's spectrum,
we find the $\chi^2_{\rm red}$ in the left bottom panel of Fig.~\ref{f-nk} and,
combining the two techniques, we get the one in the right bottom panel. In the
latter case, I have assumed $\tilde{i}=i$, since the continuum-fitting method needs
in any case $i$ as input parameter. However, we can note that the measurements
from the iron line profile and the continuum-fitting method of the spin parameter
assuming (incorrectly) that the object is a Kerr BH are not completely consistent.
The analysis of the K$\alpha$ iron line would suggest $a_* = 0.47 \pm 0.04$.
The continuum-fitting method would measure $a_* = 0.68^{+0.03}_{-0.07}$.

\section{Summary and conclusions \label{s-c}}

Astrophysical BH candidates are thought to be the Kerr BHs predicted by 
General Relativity, but the actual nature of these objects has still to be verified. 
In order to confirm the Kerr BH hypothesis, we have to 
observe strong gravity features and check that they are consistent with the 
predictions of General Relativity. It turns out that one can do it by extending 
current techniques used to estimate (under the assumption of Kerr background)
the spin of these objects. The continuum-fitting method and the analysis of 
the broad K$\alpha$ iron line are the two most robust approaches and
they can already provide constraints on possible deviations from the
Kerr geometry with the available X-ray data. These two features make
the continuum-fitting method and the K$\alpha$ iron line analysis particularly
interesting in comparison with other approaches discussed in the
literature and possible only after the advent of future experiments.

In the present paper, I focused the attention on the analysis of the broad
K$\alpha$ iron line, which is quite commonly seen in the X-ray spectrum 
of both stellar-mass and super-massive BH candidates and it is interpreted
as the fluorescent lines produced by X-ray irradiation of the cold gas
in the accretion disk. Here, I extended previous work in the literature. 
There is a significant correlation between spin parameter $a_*$, 
deformation parameter $\epsilon_3$, and the disk's inclination angle $i$. 
In the Kerr metric, the Keplerian velocity profile at relatively large radii is 
quite independent of the exact value of the spin parameter and therefore
the determinations of $a_*$ and $i$ from the shape of the line are 
not very much correlated. In the case of a non-Kerr space-time, the deformation
parameter can affect the Keplerian velocity profile, with the result
that $a_*$, $\epsilon_3$, and $i$ are not independent any more
and the constraint on possible deviations from the Kerr geometry
becomes weaker. Despite that, the analysis of the broad K$\alpha$
iron line can potentially provide stronger bounds on $\epsilon_3$
than the study of the thermal component of the accretion disk, as
can be easily seen by checking Tabs.~\ref{tab1} and \ref{tab2}
(the continuum-fitting method cannot put any bound in the region $-0.25 < a_* < 1.25$ 
and $-5 < \epsilon_3 < 15$ if the object does not rotate very fast).


\begin{acknowledgments}
I wish to thank Matteo Guainazzi, for useful comments and suggestions,
and Zilong Li, for help in the preparation of the manuscript. 
This work was supported by the Thousand Young Talents Program, 
Fudan University, and the Humboldt Foundation.
\end{acknowledgments}


\end{document}